\documentclass[11pt,final]{article}

\usepackage[utf8]{inputenc}
\usepackage{amsmath, amssymb, amsfonts, mathrsfs}
\usepackage{graphicx}
\usepackage{hyperref}
\usepackage{authblk}
\usepackage{geometry}
\usepackage{color}
\usepackage{showlabels}
\usepackage{svg}          
\usepackage{todonotes}
\usepackage{endnotes}
\usepackage{float}		  

\usepackage{caption}
\usepackage{subcaption}
\usepackage{amsthm}
\usepackage{tikz}
\usetikzlibrary{decorations.pathreplacing,arrows.meta,calc}
\theoremstyle{remark}
\newtheorem{remark}{Remark}

\usepackage{cite}
\title{Using stochastic thermodynamics with internal variables to capture orientational spreading in cell populations undergoing cyclic stretch}
\author[1]{Rohan Abeyaratne}
\author[2]{Sanjay Dharmaravan}
\author[3]{Giuseppe Saccomandi}
\author[4]{Giuseppe Tomassetti}
\affil[1]{Mechanical Engineering Department, Massachusetts Institute of Technology}
\affil[2]{Department of Mathematics and Statistics, Bucknell University}
\affil[3]{Department of Engineering, University of Perugia}
\affil[4]{Department of Industrial, Electronic, and Mechanical Engineering, Roma Tre University}

\date{}

\begin{document}
	
	\maketitle
	
	\begin{abstract}
		We revisit the modeling framework introduced in  [N. Loy and L. Preziosi: Bull. Math. Bio., 85, 2023] to describe the dynamics of cell orientation under cyclic stretch. We propose a reformulation based on the principles of Stochastic Thermodynamics with Internal Variables introduced in [T.~Leadbetter, P.~Purohit, and C.~Reina: PNAS Nexus, 2, 2023]. This approach allows us to describe not only the evolution of the orientation distribution, but also the observed spreading phenomenon. \color{black}The insight provided by our model reveals an interesting phenomenon, which we call two-stage reorientation: when cells begin aligned with an energy maximum, their orientations  spread before concentrating at the energy minimum. This theoretical prediction suggests a new experiment to test this modeling framework.\color{black}
	\end{abstract}

	
	
	\tableofcontents
	
	\section{Introduction}
	In the cardiovascular system, the endothelial cells that line the arteries experience cyclic stretching with each heartbeat due to pulsatile blood flow. Understanding how these cells respond to such \color{black}mechanical signals is believed to be crucial in studying vascular remodeling, hypertension, and atherosclerosis~\cite{jufri2015mechanical}.
	
\begin{figure}[H]
	\centering
	\includegraphics[width=0.5\linewidth]{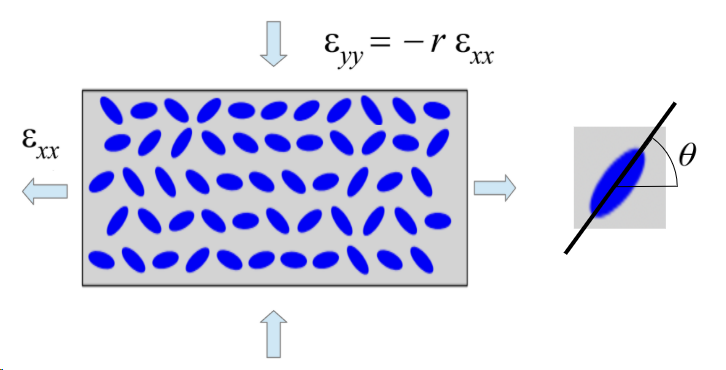}
	\caption{Sketch of the experimental setup.}\label{fig:setup}
\end{figure}

	In vivo observations in blood vessels show that the cells that form the walls of the arteries assume specific orientations, depending on their location~\cite{buck1979longitudinal}. Furthermore, in vitro experiments demonstrate that when a cell population is seeded on a cyclically stretched substrate, these cells reorient themselves in specific directions with respect to the stretching direction, and that this behavior is common to several types of cells~\cite{buck1980reorientation}. More precisely, in response to an imposed in-plane \emph{cyclic biaxial strain}: 
	\begin{equation}\label{strain}
		\underline{\boldsymbol\varepsilon}(t)=\operatorname{diag}(\varepsilon_{xx}(t),\varepsilon_{yy}(t)),
	\end{equation}
	with 
	\begin{equation}
		\varepsilon_{xx}(t)=(1+\cos\omega t)\varepsilon_{\rm avg},\qquad \varepsilon_{yy}(t)=-r\varepsilon_{xx}(t),\qquad 0\le r\le 1,
	\end{equation}
	(see Fig.~\ref{fig:setup}) each cell develops stress fibers that link to the substratum via focal adhesions along a common orientation $\theta$ with the the stretching direction. Reorientation takes place also under static conditions ($\omega = 0$), however, cells tend to align parallel to the $x$-axis, that is, along the direction of maximum stretch \cite{eastwoodEffectPreciseMechanical1998,collinsworthOrientationLengthMammalian2000}. In contrast, for $\omega\ge 1{\rm Hz}$, the common orientation attained by the cells departs from the $x$-axis by an amount that depends on the \emph{biaxiality ratio} $r$ \cite{faustCyclicStressMHz2011,livne2014cell}. Strains of $2{-}4\%$ are sufficient to trigger the reorientation process, although it may take several hours to complete. In contrast, for an applied strain of the order of $10\%$, full reorientation occurs in a time \color{black}that ranges from several minutes to a few hours\color{black} 
\cite{faustCyclicStressMHz2011,livne2014cell,maoCriticalFrequencyCritical2021}.

	Livne et al.~\cite{livne2014cell} had the intuition that cells reorient themselves to minimize an energy that depends on their orientation~\cite{de2007dynamics, livne2014cell}. According to this point of view, the evolution of the orientation angle $\theta(t)$ is governed by the differential equation	
	\begin{equation}\label{eq:deterministic}
		\frac{{\rm d}\theta}{{\rm d}t}=\Omega(\theta),
	\end{equation}
	where 
	\begin{equation}\label{Omega}
		\Omega (\theta)=- \frac{K\varepsilon^2}\eta\mathscr{U}'(\theta) ,
	\end{equation}
	is the \emph{reorientation rate}. Here $K$ is the \emph{effective stiffness} ($[K]=\text{energy})$;  for $\langle\varepsilon_{xx}^2\rangle$ the average of $\varepsilon_{xx}^2(t)$ over one period of oscillation, $\varepsilon=\sqrt{\langle\varepsilon_{xx}^2\rangle}=\sqrt{3/2}\,\varepsilon_{\rm avg}$ is the \emph{\color{black}effective average strain} \color{black} (dimensionless); $\eta$ is the \emph{viscosity constant} ($[\eta]=\text{energy}\cdot\text{time}$), and  $\mathscr{U}'(\theta)$ the derivative of the \emph{normalized energy} \color{black}$\mathscr{U}(\theta)$  (dimensionless) which takes into account that cells respond differently when stretched parallel or perpendicular with respect to the direction of their stress fibers~\cite{de2007dynamics}. The appearance of the average strain instead of the instantaneous strain in \eqref{Omega} is motivated by the fact that reorganization of the cells occurs through internal processes such as focal adhesion remodeling, stress fiber rearrangement, and cytoskeletal remodeling, take place on large times timescales compared to the period of a single oscillation.

	It can be observed that the spreading in the histograms obtained from experiments on cell orientation is more pronounced for small small strains (see for instance \cite{barronEffectPhysiologicalCyclic2007,livne2014cell,maoCriticalFrequencyCritical2021}). This observation motivated Loy and Preziosi~\cite{loy2023statistical} to account for randomness by considering the \emph{Langevin equation}:
	\begin{equation}\label{Langevin}
		{\rm d} \theta=\Omega(\theta)\,\mathrm{d} t+\sqrt{2D}\, {\rm d}W(t),
	\end{equation}
	where $D$ is the \emph{angular diffusion coefficient} ($[D]$=1/time) and $W(t)$ is the \emph{Wiener process}  ($[dW]=\sqrt{\rm time})$. Then, the statistical information about the orientation of the cells at time $t$ is described by a probability density $p(\theta,t)$ obeying the Fokker--Planck equation \cite{riskenFokkerPlanckEquationMethods1996}:
	\begin{equation}\label{fp}
		\frac{\partial}{\partial t}p(\theta,t)+\frac{\partial}{\partial\theta}( p(\theta,t)\Omega(\theta))=D  \frac{\partial^2}{\partial \theta^2}p(\theta,t),
	\end{equation}
	which must be solved for $t\ge 0$ and $\theta\in (0,\pi)$, with periodic boundary conditions:
	\begin{equation}
		p(0,t)=p(\pi,t).
	\end{equation}
	\noindent Introducing the \emph{pseudo-temperature} $\tau^2=\eta D/K$, the diffusion coefficient can be written as
	\begin{equation}\label{diffusion}
		D=\frac{K}{\eta}\tau^2.
	\end{equation}
	When $D$ vanishes, the Fokker--Planck equation reduces to the standard transport equation, which causes the probability distribution to concentrate in the proximity of the minimum of $\mathscr{U}$, as can be seen using the method of characteristics. For $D>0$, the diffusive term causes a spreading of the probability distribution. In particular, as shown in~\cite{loy2023statistical}, the probability distribution converges to an equilibrium
	\begin{equation}\label{equilibrium}
		p(\theta,t)\to p_\infty(\theta)\propto \exp \left(-\left(\frac{\varepsilon}\tau\right)^2{\mathscr{U}(\theta)}\right)\qquad \text{as }t\to\infty.
	\end{equation}
	The peaks of the equilibrium distribution $p_\infty(\theta)$ are located at the minima of the normalized energy \color{black} $\mathscr{U}(\theta)$, with a spreading proportional to $(\varepsilon/\tau)^2$. This result confirms that greater strains  lead to more pronounced peaks in the equilibrium distribution. 

While the Fokker--Planck framework allows straightforward analysis of stationary states, studying the time-dependent behavior of its solutions is considerably more involved. In this paper,  we propose a two-dimensional dynamical system that effectively captures the main features of the solutions to the Fokker--Planck equation \eqref{fp}, and is more amenable to qualitative analysis.\color{black}

After a brief recollection of known results in Section \ref{subsec:back}, we introduce in Section~\ref{subsec:vonmises}  a two-parameter family 
\begin{equation}\label{eq:family}
    \mathscr A=\{{\tilde p}(\cdot|\mu, \kappa):\mu\in[0,\pi), \kappa\ge 0\}
\end{equation} of \emph{approximating probability distributions} of the form
 \begin{equation}\label{p-tilda00}
		{\tilde p}(\theta|\mu,\kappa)=2 f_{\rm vM}\left(2\theta| 2\mu, \frac \kappa 2 \right),\qquad \theta\in[0,\pi),
	\end{equation}
where
\begin{equation}\label{vm1}
		f_{\rm vM}(\phi|\mu,\kappa)=\frac{\exp [\kappa \cos (\phi-\mu)]}{2 \pi I_0(\kappa)},\qquad \phi\in[0,2\pi)
	\end{equation}
with $I_0$ the \emph{modified Bessel function of the first kind} of order $0$, is the standard \emph{von Mises probability distribution} \cite{MardiaJupp2000}, which is defined on the entire unit circle.
    
    Each member of $\mathscr A$ is $\pi$-periodic with respect to the angle $\theta$ and exhibits a single peak located at $\mu$, with the sharpness of the peak controlled by $\kappa$, with higher values of $\kappa$ corresponding to sharper peaks (see Fig.~\ref{fig:vM} in Sec.~\ref{sec:model})\color{black}. In particular, if $\kappa\to\infty$ then ${\tilde p}(\theta|\mu,\kappa)$ converges, in the sense of distributions, to the Dirac delta supported on $\mu$; if $\kappa\to 0$, ${\tilde p}(\theta|\mu,\kappa)$ converges to the uniform distribution. In this paper, we refer to $\mu$ and $\kappa$, respectively, as the \emph{mean orientation} and the \emph{order parameter} (our terminology is different from the standard convention in Circular Statistics \cite{MardiaJupp2000} where $\mu$ is referred to as the \color{black} mode, and \color{black}$\kappa$ as the concentration).
    
    By specializing to the Fokker--Planck equation \eqref{fp} the general framework proposed in recent work by Leadbetter et al.~\cite{reina2023statistical} we obtain \color{black}
    in Section~\ref{subsec:ODE} the following two-dimensional dynamical system (\emph{cf.} \eqref{ODE-bis}):
	\begin{equation}\label{ODE}
		\begin{aligned}
			&\frac \eta K \frac{{\rm d}\mu}{{\rm d}t}=\varepsilon^2f(\mu,\kappa),\\ 
			&\frac \eta K \frac{{\rm d}\kappa}{{\rm d}t}=\varepsilon^2 g(\mu,\kappa)- \tau^2 h(\kappa).
		\end{aligned}
	\end{equation}
	For the particular case at hand, \color{black}$f(\mu,\kappa)$, $g(\mu,\kappa)$, and $h(\kappa)$ can be expressed in terms of modified Bessel functions (\emph{cf.} \eqref{eq:def-fgh}). 
    The functions $f$ and $g$ capture the drift features, and  $h$, the diffusive feature of \eqref{fp}. Solving \eqref{ODE}, we obtain a pair $(\mu(t),\kappa(t))$ of functions, from which our approximation is constructed by plugging these functions into \eqref{p-tilda00}:\color{black} \begin{equation}\label{appr}
		p(\theta,t)\simeq {\tilde p}(\theta|\mu(t),\kappa(t)),\qquad\theta\in(0,\pi),\qquad t\ge 0.
	\end{equation}
The main advantage of our approach is the possibility to perform a qualitative phase-plane analysis of the dynamical system \eqref{ODE}. In Section~\ref{sec:qualitative}, we examine the phase space of the system  in the case when the normalized energy \color{black}$\mathscr U(\theta)$ has a single minumum at $\pi/2$. We verify that the orientation angle $\mu(t)$ asymptotically converges to the energy minimum, while the order parameter $\kappa(t)$ tends toward an equilibrium value $\kappa_*$, at which the drift term and the diffusion term in $\eqref{ODE}_2$ are in balance:\color{black}
\begin{equation}\label{limit}
	\varepsilon^2 g(\pi/2,\kappa_*) = \tau^2 h(\kappa_*).
\end{equation}
Based on the aforementioned properties of the functions $g(\pi/2,\kappa)$ and $h(\kappa)$, it is observed from \eqref{limit} that, as the effective average strain $\varepsilon$ decreases, the importance of the transport term with respect to the diffusion term is reduced, which leads to an increase of the spreading $\kappa_*$ of our approximate probability distribution, in accordance with the insights from~\cite{loy2023statistical}. Notably, since the functions $g$ and $h$ are given explicitly, we are able to derive asymptotic estimates confirming that $\kappa_* = \mathcal{O}(\tau^2 / \varepsilon^2)$. 
    
	 In Section~\ref{sec:test} we solve numerically the Fokker--Planck equation on a finite time interval, and for each time $t$ in that interval we find $\overline\mu(t)$ and $\overline\kappa(t)$ 
    such that $\tilde p(\theta|\overline\mu(t),\overline\kappa(t))$ minimizes, within the family $\mathscr A$ defined in \eqref{eq:family}, the loss of information with respect to $p(\theta,t)$, quantified by the \emph{Kullbach--Leibler divergence}: 
    \begin{equation}
		D_{\mathrm{KL}}(p(\cdot,t)\| \tilde p(\cdot|\mu,\kappa))=\int_0^\pi p(\theta,t) \log \frac{p(\theta,t)}{\tilde{p}(\theta|\mu,\kappa)} d \theta.
	\end{equation}
    We then compare  the functions $\overline\mu(t)$ and $\overline\kappa(t)$ with the functions $\mu(t)$ and $\kappa(t)$ obtained by solving \eqref{ODE}, and we find a fairly good agreement. 

In our view, an advantage of the reduced system \eqref{ODE} over the full Fokker--Planck equation lies in its ability to capture the transient dynamics through the solution of an ordinary differential equation (ODE) system.  We exploit this feature in Section~\ref{sec:experiments}, where we compare the transient response predicted by our model with both the experimental and theoretical results presented in \cite{maoCriticalFrequencyCritical2021}, observing good agreement. We give additional remarks in Section~\ref{sec:conclusions}.\color{black}

	\section{Background}\label{sec:background}
	\subsection{Cell reorientation under cyclic strain}\label{subsec:back}
	Livne et al. postulate the following deterministic \color{black}evolution equation for the orientation angle $\theta$:
	\begin{equation}\label{deterministic}
		\eta\frac{{\rm d}\theta}{{\rm d}t}=-\frac{\partial\mathcal E}{\partial\theta}(\varepsilon,\theta)=-K\varepsilon^2\mathscr U'(\theta),
	\end{equation}
	which is equivalent to \eqref{Omega}.  According to \eqref{deterministic}, the orientation angle $\theta(t)$ evolves so as to minimize a strain energy where the time average $\varepsilon^2$ over one period appears in place of the squared instantaneous strain $\varepsilon^2_{xx}(t)$. This assumption reflects the hypothesis that cellular reorientation is governed by slow internal processes (e.g., cytoskeletal remodeling, focal adhesion turnover) that are unable to track fast fluctuations in the applied strain, and therefore respond to its average over time.
        The underlying hypothesis is that the characteristic time $\eta/(\varepsilon^2K)$, which governs cell reorientation is much larger than $1/\omega$. In other words, cells are too slow to respond to the instantaneous stretch at every moment, so that the effective mechanical signal that they sense is the average $\varepsilon$ of the cyclic deformation over one cycle.

	     Following \cite{loy2023statistical}, we assume that the strain energy of the cell depends 
on the angle $\theta$  between the stress fibers and the $x$-axis. and on the strain $\varepsilon_{xx}$ according to the formula:
	\begin{equation}\label{Uloy}
\begin{aligned}
    	\mathscr{E}(\varepsilon_{xx},\theta)=\frac{\varepsilon_{xx}^2}{2}&\left\{K_\parallel[(r+1) \cos 2 \theta+1-r]^2\right.\\&\left.+{K_{\perp}}[(r+1) \cos 2 \theta-1+r]^2+{K_s}(r+1)^2\left(1-\cos ^2 2 \theta\right)\right\}.
    	\end{aligned}
    \end{equation}\color{black}
The material constant $K_\parallel$ \color{black}  quantifies the resistance to stretching along the stress fibers. The material constants $K_\perp$ and $K_s$ characterize, respectively, the resistance to stretching perpendicular to the stress fibers and the resistance to shear deformation, while $0\le r\le 1$ is the \emph{biaxiality ratio} of the imposed strain. On setting
	\begin{equation}\label{k1k2}
		k = \frac{1-r}{1+r} \frac{\displaystyle 1-\frac{K_\perp}{K}}{\displaystyle 1+\frac{K_\perp}{K}-\frac{K_s}{K}},\qquad \bar k=(r+1)^2\left(1+\frac{K_{\perp}}{K_\parallel}-\frac {K_s}{K_\parallel}\right),
	\end{equation}
	and writing
	\begin{equation}
	K=|\bar k|K_\parallel,
	\end{equation}
	we obtain $\mathscr E(\varepsilon_{xx},\theta)=K\varepsilon^2\mathscr U(\theta)$,
where the normalized energy $\mathscr U(\theta)$ has the form
	\begin{equation}\label{U}
		\mathscr{U}(\theta) = \pm\frac 1 2 \left(\cos 2\theta + k\right)^2 + c,\qquad \pm=\operatorname{sign}(\bar k),
	\end{equation}
with $c$ is a constant that does not depend on $\theta$, and that does not affect the reorientation rate:
	\begin{equation}\label{mirror}
		\Omega(\theta,t)=\pm 2\frac{K\varepsilon^2}{\eta}  (\cos 2\theta+k)\sin 2\theta.
	\end{equation}
\color{black}	
The normalized energy \color{black}$\mathscr U(\theta)$  has the following symmetry properties:
\begin{equation}\label{symmetry}
	\mathscr U(\theta)=\mathscr U(\theta+\pi),\qquad \mathscr U(\theta)=\mathscr U(-\theta),\qquad \mathscr U(\theta)=\mathscr U(\pi-\theta).
\end{equation}
The first equation in \eqref{symmetry} is consistent with the fact that stress fibers lack polarity, so that the angles $\theta$ and $\theta+\pi$ are energetically equivalent; the second and third equations are in agreement with the symmetry of the system with respect to the $x-$ and $y-$axes.

From \eqref{U} we can verify that the following types of energy landscapes are possible for $\theta$ in the interval $[0,\pi)$:
	\begin{itemize}
		\item [(a)]  if $\bar k>0$ and $-1<k<1$\color{black}, then $\mathscr U(\theta)$ has two disjoint \color{black}absolute minima (and no other local minimum), located at mirror-symmetric positions with respect to $\pi/2$, located in the interior of the interval $[0,\pi]$;\color{black}
		\item [(b)] if $\bar k<0$ and $-1<k<1$, then $\mathscr U(\theta)$ has one absolute minimum at $\theta=0$ and a local minimum at $\theta=\pi/2$; 
		\item [(c)]  if $\bar k>0$ and $k\ge 1$ or $\bar k<0$ and $k\le -1$, then $\mathscr U(\theta)$ has a minimum at $\theta=\pi/2$;
		\item [(d)] if $\bar k>0$ and $k\le-1$ or $\bar k<0$ and $k\ge 1$, then $\mathscr U(\theta)$ has a minimum at $\theta=0$.
	\end{itemize}
	Figure~\ref{fig:energy-landscapes} below illustrates the energy landscape of the function for $\bar k=1$ and for several representative values of $k\ge 0$.
\begin{figure}[htbp!]
	\centering
		\includegraphics[width=0.7\textwidth]{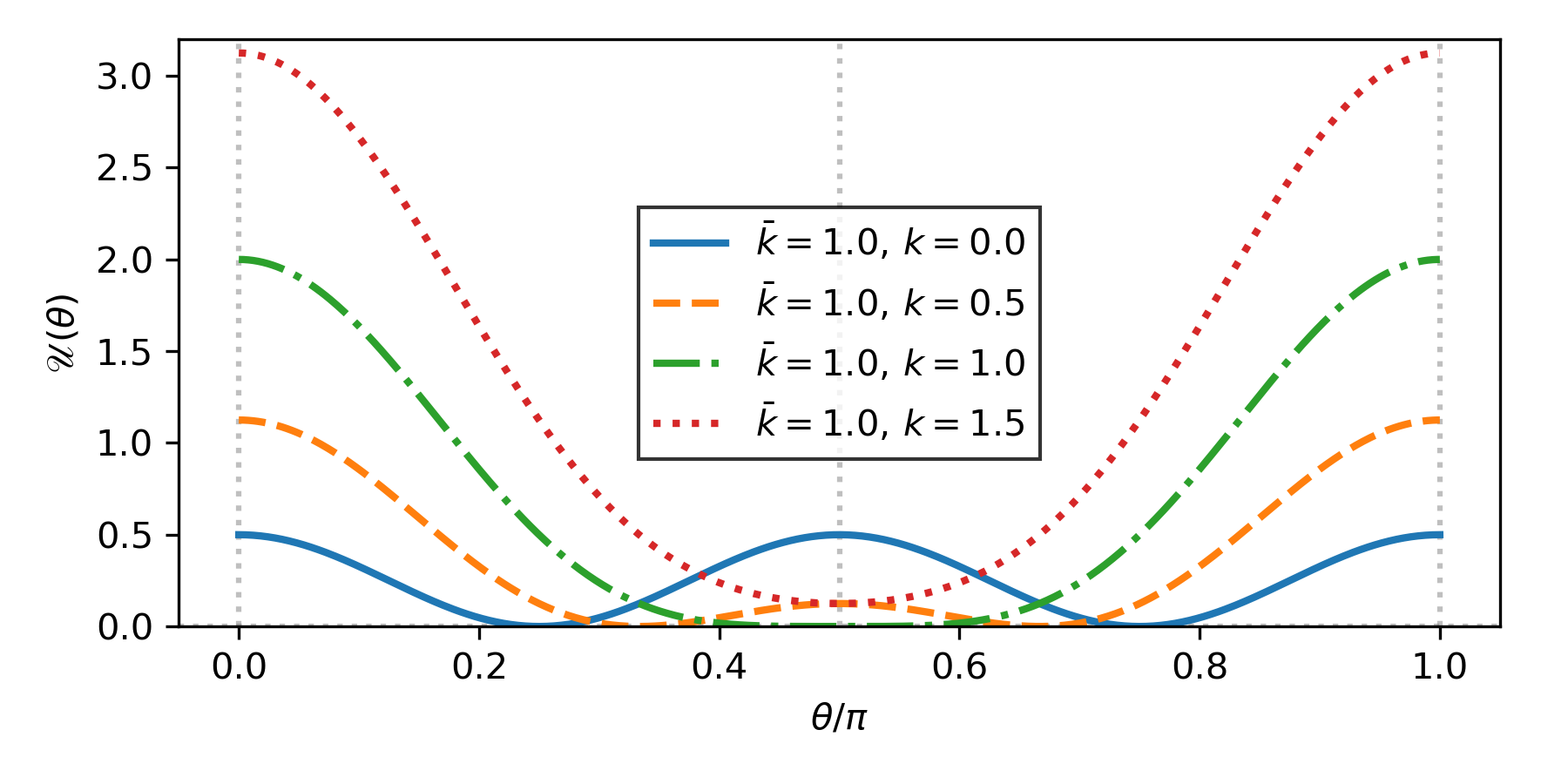}
	\caption{Energy landscapes for $\bar k=1$ and $k\ge 1$.}
	\label{fig:energy-landscapes}
\end{figure}
The energy landscape for $k<0$ can be recovered from the  profile with opposite $k$ through a shift by $\pi/2$, since $\cos(2\theta)+k=-(\cos(2(\pi/2-\theta)-k))$, as a consequence of the identity $\cos\phi=-\cos(\pi-\phi)$. The energy landscape with $\bar k=-1$ is obtained from that with $\bar k=+1$ by reflection across the horizontal axis.\color{black}

\begin{remark}[Nonlinear elasticity]
\color{black}Experiments are actually carried out for ranges of strains up to $30\%$. At these levels of strain, the use of linear elasticity is questionable. For a generalization to the nonlinear setting we refer to \cite{abeyaratneElementaryModelFocal2022,lucci2021nonlinear}.
\end{remark}

	\subsection{An approximation scheme for the Fokker--Planck equation}
	\label{subsec:lead}
	The deterministic evolution equation \eqref{deterministic} is consistent with the experimental evidence that the orientation angle approaches asymptotically the minimum of $\mathscr U(\theta)$, and that the time for reorientation process scales like the reciprocal of the square of the applied strain. However, it does not account for the spreading observed in experimental histograms reporting the distribution of cell orientation (see for instance \cite{barronEffectPhysiologicalCyclic2007,livne2014cell,maoCriticalFrequencyCritical2021}).

 \color{black}In \cite{loy2023statistical}, Loy and Preziosi introduced the Langevin equation \eqref{Langevin} as a  generalization of \eqref{eq:deterministic} to the non-deterministic case, taking the normalized energy $\mathscr U(\theta)$ as in \eqref{Uloy}. They proved the convergence of solutions of the Fokker-Planck equation to the stationary distribution $p_{\infty}(\theta)$ (see \eqref{equilibrium}), confirming that their stochastic model captures the increase in spreading that accompanies a decrease in strain, as observed in equilibrium histograms. Furthermore, by performing Monte Carlo simulations of the Langevin equation, they demonstrated that the model matches the experimentally measured statistics of cell orientations over time..\color{black}

Recently, Leadbetter et al. \cite{reina2023statistical} devised an approach to approximate the solutions of  a \color{black} \emph{prototype Fokker--Planck equation}
	\begin{equation}
		\frac{\partial p}{\partial t}=\mathcal Lp.
	\end{equation}
	In their general setting, $p=p(x,t)$ is a time-dependent probability distribution with $x$ in some domain $X$ \color{black}\color{black}. Their approximation has the form
	\begin{equation}\label{p1}
		p(x,t)\simeq \hat p(x|\boldsymbol{\alpha}(t)),\qquad x\in X,
	\end{equation}
	where the vector $\boldsymbol\alpha(t)=(\alpha_1(t),\ldots,\alpha_N(t))$ consists of $N$ scalar \emph{internal state variables}. One of their main results is that the best approximation of the form \eqref{p1} is obtained if the internal variables obey the following system of evolution equations (see \cite[Eq. 7]{reina2023statistical})
	\begin{equation}\label{ODEreina}
		\sum_{j=1}^N \left( \frac{\partial \hat{s}}{\partial \alpha_i}, \frac{\partial \hat{s}}{\partial \alpha_j} \right)_{\hat p} \dot{\alpha_j} = - k_B \left( \mathcal{L}^{\dagger} \frac{\partial \hat{s}}{\partial \alpha_i}, 1 \right)_{\hat p}, \qquad i = 1, \ldots, N,
	\end{equation}
	where
	\begin{equation}\label{entropy}
		\hat{s}(x|\boldsymbol{\alpha})=-k_B\log (\hat{p}(x| \boldsymbol{\alpha}))
	\end{equation}
	with $k_B$ the Boltzmann constant ([$k_B$]=energy/temperature), is the approximate entropy, \color{black}$\mathcal L^\dagger$ is the adjoint of the \emph{evolution operator} $\mathcal L$ and, given two functions $\varphi_1(x|\boldsymbol\alpha)$ and $\varphi_2(x|\boldsymbol\alpha)$, corresponding to two observable quantities $\Phi_1(\boldsymbol{\alpha})$ and $\Phi_2(\boldsymbol{\alpha})$, 
	\begin{equation}\label{scalar}
		(\varphi_1,\varphi_2)_{\hat p}\equiv (\varphi_1,\varphi_2)_{\hat p}[\boldsymbol\alpha]:= \int_X \varphi_1(x|\boldsymbol{\alpha}) \varphi_2(x|\boldsymbol\alpha)
		\hat p(x|\boldsymbol{\alpha}) dx
	\end{equation}
	is the expectation of $\Phi_1(\boldsymbol{\alpha}) \Phi_2(\boldsymbol{\alpha})$ under the probability distribution $\hat p(x|\boldsymbol\alpha)$. The application of the framework mentioned above in the present work is elaborated in the next section\color{black}, where we take $X=(0,\pi)$.\color{black}

		\begin{remark}[Symmetry-preserving dynamics]
	If the initial condition for the Fokker--Planck equation is the uniform probability distribution, then, in view of the symmetry properties \eqref{symmetry} of the dimensionless energy $\mathscr U(\theta)$, the solution is mirror--symmetric with respect to $\pi/2$. In particular, at each time $t$ both the drift and the diffusion terms would vanish at $0$ and $\pi/2$. Therefore, one may obtain the solution by working in the interval $(0,\pi/2)$ with zero-flux conditions at $0$ and $\pi/2$.
	\end{remark}

	\section{The reduced model}\label{sec:model}

	\subsection{Approximation through the von Mises probability distribution}\label{subsec:vonmises}
	In all of the cases discussed in the previous subsection, except for the first one, the equilibrium probability distribution $p_\infty(\theta)$ defined in \eqref{equilibrium} has only one peak. Thus, with the exception of case (a), we may approximate the probability distribution $p(\theta,t)$ at finite time using a unimodal probability distribution that contains information about the location of the peak and its spreading. To this aim, a good candidate appears to be the \emph{von Mises} probability distribution introduced in \eqref{vm1}.
    
	Among all  circular probability distributions $f:[0,2\pi)\to\mathbb R$, \color{black}the  von Mises distribution maximizes the Shannon  information entropy  \cite{shannonMathematicalTheoryCommunication1948} \color{black} $-\int_0^{2 \pi} f(\phi) \log f(\phi) d \phi$ under the constraint that the trigonometric moments $\langle \cos\phi\rangle_f=\int_0^{2\pi}\cos\phi\, f(\phi)d\phi$ and $\langle \sin\phi\rangle_f=\int_0^{2\pi}\sin\phi\, f(\phi)d\phi$ \cite{MardiaJupp2000} be fixed\color{black}; it was introduced by Richard von Mises in his investigations relating to the analysis of errors in data with circular distributions \cite[Eq. 12]{vonMises1918reprint}, and it is the  analogue of the Gaussian distribution when the underlying data is distributed in the unit circle, with $1/ {\sqrt{\kappa}}$ being the analogue of the variance. 
	
\begin{figure}[H]
\centering
\includegraphics[width=0.5\linewidth]{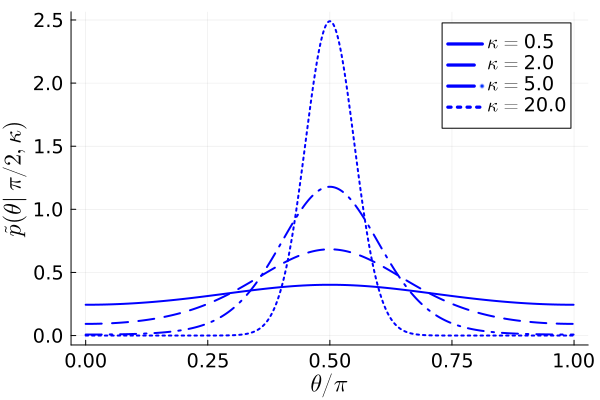}	
\caption{Graph of the rescaled von Mises distribution for $\mu=\pi/2$ and $\kappa$ ranging from $0.5$ to $20$.}
\label{fig:vM}
\end{figure}

	The standard von Mises distribution is defined on the entire unit circle. In the problem at hand, the angular variable $\theta$ only spans the interval $[0,\pi)$. We therefore perform a change of variables by defining the \emph{$\pi$-periodic rescaled von Mises distribution} as in \eqref{vm1}, that is,
	\begin{equation}\label{p-tilda}
		{\tilde p}(\theta\mid \mu,\kappa)=\frac{\exp \left[\frac \kappa 2 \cos (2(\theta-\mu))\right]}{ \pi I_0\left(\frac \kappa 2\right)} , \quad \theta\in[0,\pi),
	\end{equation}
and we seek time dependent coefficients $\mu(t)$ and $\kappa(t)$ so that the probability distribution ${\tilde p}(\theta|\mu(t),\kappa(t))$ 
	approximates the solution $p(\theta,t)$ of the Fokker--Planck equation \eqref{fp} in the interval $(0,\pi)$ with periodic boundary conditions. Plots of $\tilde p(\theta|\mu,\kappa)$ are given in Fig.~\ref{fig:vM}. For $\kappa$ small the distribution becomes flat. For $\kappa$ large, the distribution concentrates near $\pi/2$.
To choose the best $\mu(t)$ and $\kappa(t)$, we rely on the approximation scheme proposed in \cite{reina2023statistical}, which we have described in the previous section. Here we take $X$ to be the space of periodic twice-differentiable functions on $(0,\pi)$. In view of \eqref{p-tilda} and mimicking \eqref{entropy}, we define the approximate entropy as:
\color{black}\begin{equation}\label{eq:entropy}
	\tilde s(\theta|\mu,\kappa)=-K\log \tilde p(\theta|\mu,\kappa).
	\end{equation}
	System \eqref{ODEreina} specializes to 
	\begin{equation}\label{ODE2}%
		\begin{aligned}
			&\left\|\frac{\partial \tilde{s}}{\partial \mu}\right\|_{\tilde{p}}^2 \dot{\mu} + \left( \frac{\partial \tilde{s}}{\partial \mu}, \frac{\partial \tilde{s}}{\partial \kappa}\right)_{\tilde{p}} \dot{\kappa} = - K\left( \mathcal{L}^\dagger \frac{\partial \tilde{s}}{\partial \mu}, 1 \right)_{\tilde{p}}, \\
			&\left( \frac{\partial \tilde{s}}{\partial \mu} ,\frac{\partial \tilde{s}}{\partial \kappa}\right)_{\tilde{p}} \dot{\mu} + \left\|  \frac{\partial \tilde{s}}{\partial \kappa} \right\|^2_{\tilde{p}} \dot{\kappa} = - K\left( \mathcal{L}^\dagger \frac{\partial \tilde{s}}{\partial \kappa}, 1 \right)_{\tilde{p}},
		\end{aligned} 
	\end{equation}
	where $\|\cdot\|_{\tilde p}$ is the norm associated to the scalar product $(\cdot,\cdot)_{\tilde p}$. Here, the operator $\mathcal{L}^\dagger$, defined by
	\begin{equation}
	\mathcal{L}^\dagger \psi = \Omega \frac{\partial\psi}{\partial\theta}+ D \, \frac{\partial ^2\psi}{\partial\theta^2}
	\end{equation}
is the \emph{adjoint} of the \emph{evolution operator} $\mathcal L$, defined by $\mathcal{L}p = -\frac{\partial(\Omega p)}{\partial\theta} + D \, \frac{\partial^2 p }{\partial\theta^2}$.

\begin{remark}[Non-polarity of the cells]
     We have already pointed out that the angle $\theta$ spans the interval $[0,\pi)$ because cell fibers exhibit no polarity. If the fibers were polar, then the angle $\theta$ would span the interval $[0,2\pi)$. In this case, the approximate probability distribution could modeled using the standard von Mises distribution.
\end{remark}

	\begin{remark}[\color{black}Interpretation of $\tau^2$ as the analogue of temperature]
  	The quantity $K\tau^2$ has the dimension of energy. Classically, this quantity would be written as $k_BT$, where $k_B$ is Boltzmann's constant and $T$ is temperature. In the present setting, however, fluctuations are not of thermal nature \cite{dasCellReorientationCyclically2022, shishvan2018homeostatic}. It is therefore natural for us to use $K$ in place of $k_B$ in the definition of entropy, \color{black} and to interpret $\tau^2$ as the analogue of temperature.\color{black}
        \end{remark}
	
	\subsection{The coefficients of the ODE system that governs the average angle and the order parameter}\label{subsec:ODE}
	The coefficients on the left-hand side of \eqref{ODE2} can be computed in closed form using the definition \eqref{scalar} of the symbol $(\cdot,\cdot)_{\tilde p}$ with the domain of integration taken to be $(0,\pi)$. The resulting expressions involve combinations of modified Bessel functions $I_\nu(z)$ \cite{abramowitz1964handbook}. We find
	\begin{equation}\label{example}
    \begin{aligned}
		&\footnotesize\left\|\frac{\partial \tilde{s}}{\partial \mu}\right\|^2_{\tilde{p}}=\int_0^{\pi}\left(\frac{\partial \tilde{s}(\theta|\mu,\kappa)}{\partial \mu}\right)^2\tilde p(\theta|\mu,\kappa){\rm d}\theta=4K^2\frac \kappa 2  \frac{I_1(\kappa/2)}{I_0(\kappa/2)},\\
			&\left\|\frac{\partial \tilde{s}}{\partial \kappa}\right\|_{\tilde{p}}^2=\int_0^{\pi}\left(\frac{\partial \tilde{s}(\theta|\mu,\kappa)}{\partial \kappa}\right)^2\tilde p(\theta|\mu,\kappa){\rm d}\theta=\frac{K^2}{4}\left(1-\frac{I_1^2(\kappa/2)}{I_0^2(\kappa/2)}-\frac{1}{\kappa/2} \frac{I_1(\kappa/2)}{I_0(\kappa/2)}\right),\\
			&\left(\frac{\partial \tilde{s}}{\partial \mu}, \frac{\partial \tilde{s}}{\partial \kappa}\right)_{\tilde{p}}=\int_0^{\pi}\left.\frac{\partial \tilde{s}(\theta|\mu,\kappa)}{\partial \mu}\right.\left.\frac{\partial \tilde{s}(\theta|\mu,\kappa)}{\partial \kappa}\right.\tilde p(\theta|\mu,\kappa){\rm d}\theta=0.
		\end{aligned}
	\end{equation}\color{black}
	The coefficients on the right-hand side of \eqref{ODE2} also admit a representation in terms of known functions:   
	\begin{equation}\label{eq:right-hand-sides}
		\begin{aligned}
			&-\left(\mathcal L^\dagger\frac{\partial\tilde s}{\partial\mu},1\right)_{\tilde p}
			=
			8\varepsilon^2\frac{K k_1}\eta 
			\left(
			\frac
			{\frac \kappa 2 I_1(\frac \kappa 2)-2I_2(\kappa/2)}{I_0(\kappa/2)}
			\cos(2\mu)
			+ 
			\frac
			{I_1(\kappa/2)+\frac \kappa 2 I_2(\frac \kappa 2)}
			{I_0(\frac \kappa 2)} k_2 
			\right)
			\sin(2\mu),
			\\
			&-\left(\mathcal L^\dagger\frac{\partial\tilde s}{\partial\kappa},1\right)_{\tilde p}
			=
			-2\varepsilon^2\frac{K k_1}\eta
			\left(
			\frac{I_2(\kappa/2)}{\frac \kappa 2I_0(\kappa/2)}
			\cos(4\mu)
			+
			\frac
			{I_1(\frac \kappa 2)}
			{\frac \kappa 2 I_0(\frac \kappa 2)}
			k_2\cos(2\mu)\right)-\tau^2\frac{K}{\eta}\frac{ I_1(\kappa/2)}{I_0(\kappa/2)}
			.
		\end{aligned}
	\end{equation}
	Using these result, we obtain the ODE system \eqref{ODE} with
	\begin{equation}\label{eq:def-fgh}
		\begin{aligned}
			f(\mu,\kappa) 
			&= 2k_1 F_1(\kappa)
			\left(
			F_2(\kappa) \cos 2\mu +
			k_2 \right) \sin 2\mu,
			\\
			g(\mu,\kappa)&=-k_1 G(\kappa)\left(\frac{I_2(\frac\kappa 2)}{I_1(\frac\kappa 2)}(2(\cos 2\mu)^2-1)+k_2\cos2 \mu\right),\\
			h(\kappa)&=
			\frac{8 }{\frac{I_0(\frac\kappa 2)}{I_1(\frac\kappa 2)}-\frac{I_1(\frac\kappa 2)}{I_0(\frac\kappa 2)}-\frac 2 \kappa},
		\end{aligned}
	\end{equation}
	where
	\begin{equation}
		\begin{aligned}
			&F_1(\kappa)=\frac{I_2\left( \frac{\kappa}{2} \right)}{I_1\left( \frac{\kappa}{2} \right)}+\frac 1 {\frac \kappa 2},\qquad F_2(\kappa)=\frac{1-2\frac{I_2\left( \frac{\kappa}{2} \right)}{\frac \kappa 2 I_1\left( \frac{\kappa}{2} \right)}  }{\frac{ I_2\left( \frac{\kappa}{2} \right)}{I_1\left( \frac{\kappa}{2} \right)}+\frac 1 {\frac \kappa 2}},\qquad G(\kappa)=\frac{8} {\frac \kappa  2 \left(\frac{I_0\left( \frac{\kappa}{2} \right)}{I_1\left( \frac{\kappa}{2} \right)}- 
				\frac{I_1\left( \frac{\kappa}{2} \right)}{I_0\left( \frac{\kappa}{2} \right)}
				\right)- 1}=\frac{h(\kappa)}{\kappa /2}.
		\end{aligned}
	\end{equation}
    The properties of the above functions will be examined in Section \eqref{sec:qualitative}.
    \begin{remark}[Diagonal structure of the ODE system]
        It is worth emphasizing that the off diagonal entries on the left-hand side of \eqref{ODE2} vanish by $\eqref{example}_3$. \color{black}This structural simplification plays a key role in facilitating the analysis of the system.
    \end{remark}
	\begin{remark}[Closed-form of the coefficients \eqref{example}]
		The computation of integrals like \eqref{example} in terms of known functions is possible because the $\pi$-periodic von Mises distribution $\tilde{p}(\phi;\mu,\kappa)$ admits the known Fourier series
		\[
		\tilde{p}(\phi;\mu,\kappa) = \frac{1}{\pi} \left[ 1 + 2 \sum_{n=1}^\infty \frac{I_n(\kappa/2)}{I_0(\kappa/2)} \cos(2n(\phi - \mu)) \right],
		\]
and because in all calculations the quantities acted upon by the expectation operator have a finite number of harmonics. 
	\end{remark}
		
\subsection{Setting the initial conditions}\label{sec:initial_conditions}
The Fokker--Planck equation \eqref{fp} comes with an initial condition of the form
\begin{equation}
    p(\theta,0) = p_0(\theta), \quad \theta \in (0, \pi).
\end{equation}
In order to employ the reduced ODE system \eqref{ODE}, it is necessary to provide corresponding initial conditions:
\begin{equation}\label{ic}
    \mu(0) = \mu_0, \quad \kappa(0) = \kappa_0.
\end{equation}
\color{black}When the initial condition $p_0(\theta)$ for the Fokker--Planck equation is an element of the approximating family $\mathscr A$  defined in \eqref{eq:family}, there exist $\mu_0$ and $\kappa_0$ such that $p_0(\theta)=\tilde p(\theta|\mu_0,\kappa_0)$. In this, case $(\mu_0,\kappa_0)$ serve as the appropriate initial condition  for the ODE system \eqref{ODE}. For example, this holds true when $p_0(\theta)$ is the uniform probability distribution (see Remark~\ref{rem:specialCases} below). In general, however, $p_0(\theta)$ needs not belong to the family $\mathscr A$, and we need a criterion \color{black} to select the initial state $(\mu_0, \kappa_0)$ of the dynamical system \eqref{ODE} so as to best represent $p_0(\theta)$ within the family $\mathscr A$ of approximating probability distributions. 

We propose that the initial state should be selected by solving the minimization problem
	\begin{equation}\label{minimization}
		\big(\mu_0,\kappa_0\big)=\operatorname{argmin}_{\big(\bar\mu,\bar\kappa\big)}D_{\mathrm{KL}}(p_0\| {\tilde p}(\cdot ; \bar\mu, \bar\kappa)),
	\end{equation}
	where 
	\begin{equation}\label{def:DKL}
		D_{\mathrm{KL}}(p_1\| p_2)=\int_0^\pi p_1(\theta) \log \frac{p_1(\theta)}{p_2(\theta)} d \theta
	\end{equation}
	denotes the \emph{Kullback-Leibler (KL) divergence} between the probability distributions $p_1$ and $p_2$. The K--L divergence provides a measure of the loss of information when the probability distribution $p_2$ is used to approximate $p_1$ \cite{kullback1951information,kullback1959information}. Thus, 
	\begin{equation}
p_0(\theta,0)\simeq{\tilde p}(\theta;\mu_0,\kappa_0)
	\end{equation}
	represent the best information-theoretic approximation of the initial datum $p_0(\theta)$ within the family $\mathscr A$ of approximating functions defined in \eqref{eq:family}.\color{black}

We now show that the solution of the minimization problem \eqref{minimization}  is the solution of the following system:
	\begin{equation}\label{initial2}
		\hat{\bf r}(2\mu_0)\text{ is parallel to } \langle\hat{\bf r}(2\theta)\rangle_{p_0},\qquad \frac{I_1(\kappa_0/2)}{I_0(\kappa_0/2)} = \lvert \langle \hat{\bf r}(2\theta)\rangle_{p_0} \rvert.
	\end{equation}
	Here
	\begin{equation}
		\hat{\bf r}(\phi)=(\cos\phi,\sin\phi),\qquad \phi\in[0,2\pi),
	\end{equation}
	is a unit circle parametrization, and $\langle f(\theta)\rangle_{p_0}=\int_0^\pi f(\theta) p_0(\theta)d\theta$ denotes the expectation of $f(\theta)$ with respect to the probabilty distribution $p_0(\theta)$.

	Let us use the shorthand notation $D_{\rm KL}(\overline\mu,\overline\kappa)$ in place of $D_{\mathrm{KL}}(p_0\| {\tilde p}(\cdot ; \overline\mu, \overline\kappa))$. To seek the minimizer $(\mu_0,\kappa_0)$ of $D_{\rm KL}(\overline\mu,\tilde\kappa)$ with respect to $\overline\mu$ and $\overline\kappa$ we solve
	\begin{equation}\label{minima}
		\frac{\partial D_{\text{KL}}}{\partial \overline\mu}(\mu_0,\kappa_0)=0,\qquad \frac{\partial D_{\text{KL}}}{\partial \overline\kappa}(\mu_0,\kappa_0)=0.
	\end{equation}
	Differentiating $D_{\rm KL}(\overline\mu,\bar\kappa)$ with respect to \(\bar\mu\) gives
	\begin{equation}
		\frac{\partial D_{\text{KL}}}{\partial \bar\mu}= \frac{\kappa}{2} \int_0^\pi p_0(\theta)\sin(2\theta-2\bar\mu)\,d\theta.
	\end{equation}
	Thus, the first of \eqref{minima} yields
	\[
	\langle \sin(2\theta-2\mu_0) \rangle_{p_0} = 0,
	\]
	that is, the expected value of $\sin(2\theta-2\mu_0)$ with respect to the initial probability distribution $p_0(\theta)$ vanishes. By the addition formula for the sine function, 
	\begin{equation}
	\langle\sin(2\theta)\rangle_{p_0}\cos(2\mu_0)-\langle\cos(2\theta)\rangle_{p_0}\sin(2\mu_0)=0.
	\end{equation}
	The last equation is equivalent to the requirement that  the double-angle unit vector 
	\begin{equation}
	\hat{\bf r}(2\mu_0) = (\cos2\mu_0,\,\sin2\mu_0)
	\end{equation}
	associated to $\mu_0$ is aligned with the expected value, under the probability distribution $p_0$, of the double-angle vector associated to $\theta$:
	\begin{equation}\label{mu0}
		\langle \vec{\bf r}(2\theta)\rangle_{p_0} = \Bigl(\langle\cos2\theta\rangle_{p_0},\;\langle\sin2\theta\rangle_{p_0}\Bigr),
	\end{equation}
	as stated by the first of $\eqref{initial2}$. \color{black} 
	Similarly, differentiating \(D_{\text{KL}}(\overline\mu,\bar\kappa)\) with respect to \(\bar \kappa\) produces
	\begin{equation}
		\frac{\partial D_{\text{KL}}}{\partial \bar\kappa} = \frac{1}{2}\left(\langle \cos(2\theta-2\bar\mu)\rangle_{p_0} - \frac{I_1(\bar\kappa/2)}{I_0(\bar\kappa/2)}\right).
	\end{equation}
	Then, in view of \eqref{mu0}, it is not difficult to see that \eqref{minima} is equivalent to the second of \eqref{initial2}.
	\begin{remark}[Special cases]\label{rem:specialCases}
		If the initial probability distribution is uniform, that is $p_0(\theta)=1/\pi$, then $\langle \vec{\bf r}(2\theta)\rangle_{p_0}=(0,0)$. Then, system \eqref{initial2} requires $\kappa_0=0$, but does not determine $\mu_0\in[0,\pi)$. Conversely, if $p_0(\theta)$ is a Dirac delta, then $\mu_0$ coincides with the support of $p_0$, and $\kappa_0=\infty$.
	\end{remark}

\section{The qualitative behaviour of the reduced model}\label{sec:qualitative}
\begin{figure}[H]
	\centering
	\begin{subfigure}[t]{0.48\textwidth}
		\centering
		\includegraphics[width=\linewidth]{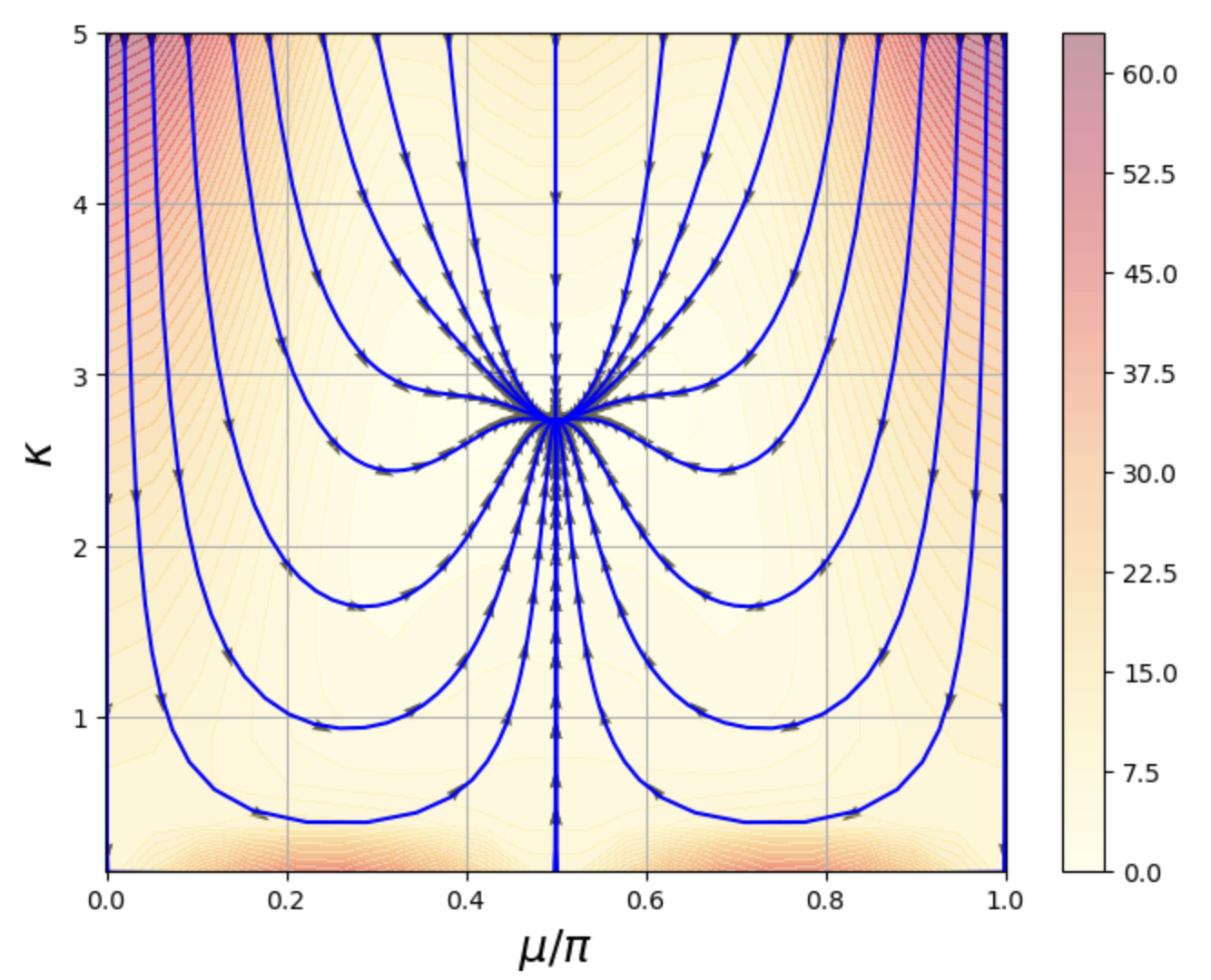}
		\caption{Trajectories of the solutions of the ODE system~\eqref{ODE-bis}. Arrows indicate the direction of the flow. Along each trajectory, arrows are spaced evenly in time; denser arrows indicate slower motion. Color intensity encodes the norm of the velocity field.}
	\end{subfigure}
	\hfill
	\begin{subfigure}[t]{0.48\textwidth}
		\centering
		\includegraphics[width=1.05\linewidth]{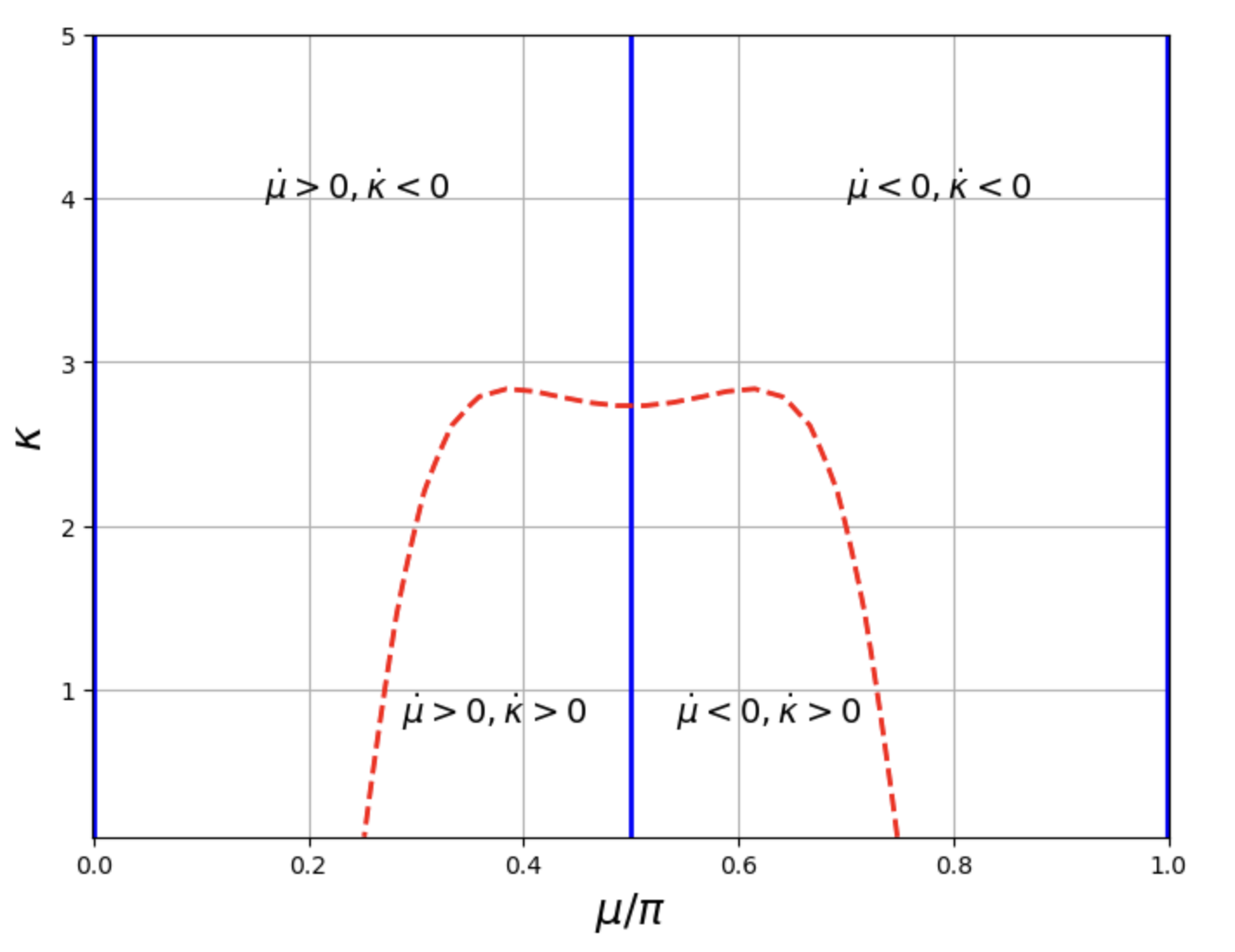}
		\caption{Zero-level sets of the right-hand sides of system~\eqref{ODE-bis}. The rate $\dot\mu$ vanishes on the blue solid curves, while the rate $\dot\kappa$ vanishes on the red, dashed curve.}
	\end{subfigure}
	\caption{Phase portrait and structure of the solution space for the ODE system~\eqref{ODE-bis}, with parameters $\tau=\varepsilon=1$, $k = 1$, $\bar k = 1$.}
	\label{fig:phaseportrait}
\def\code{phase\_portrait2.ipynb}
\end{figure}

\begin{figure}[htbp]
    \centering
        \centering
        \includegraphics[width=0.5\textwidth]{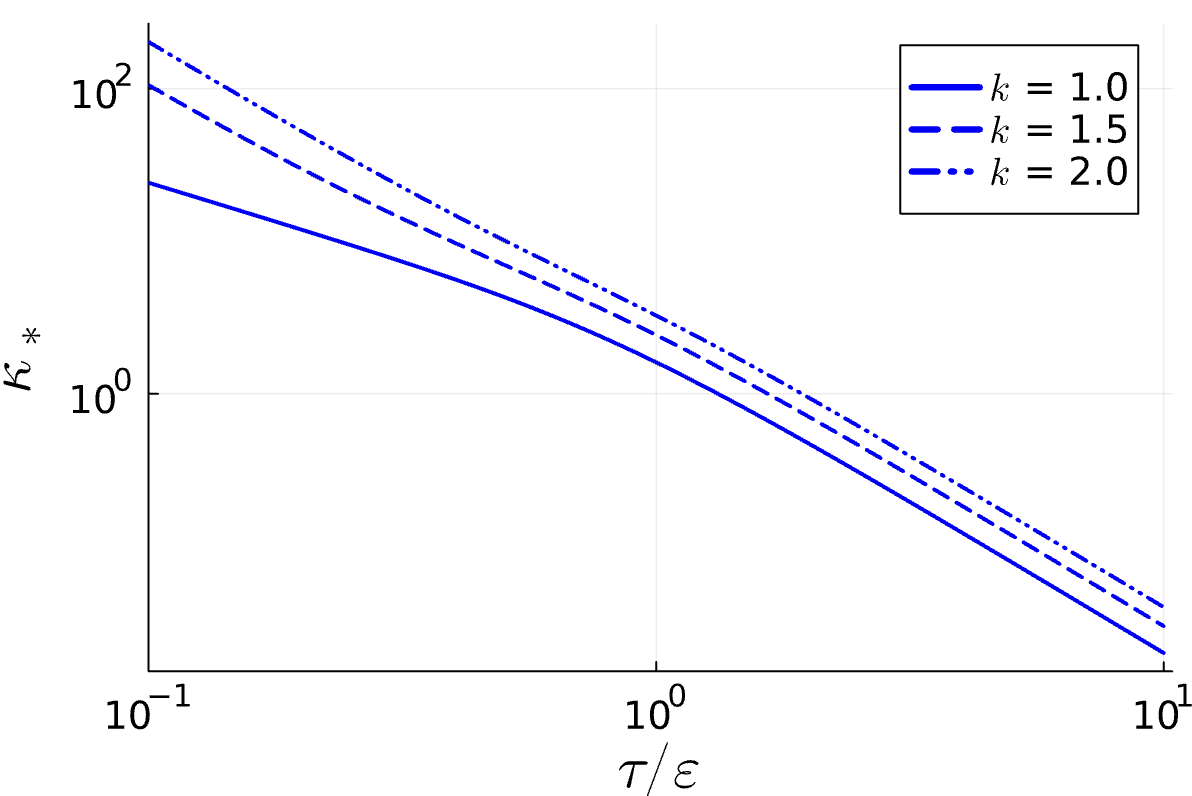}
        \caption{Dependence of $\kappa_*$ on the ratio $\tau/\varepsilon$ for different values of the constant $k$.}
        \label{fig:stationary_state_kappa}
\end{figure}

In this section we analyse the qualitative behaviour of the solutions of the dynamical system  with governing equations \eqref{ODE}, which we repeat here for convenience:
\begin{equation}\label{ODE-bis}
	\begin{aligned}
		&\frac \eta K \frac{{\rm d}\mu}{{\rm d}t}=\varepsilon^2f(\mu,\kappa),\\ 
		&\frac \eta K \frac{{\rm d}\kappa}{{\rm d}t}=\varepsilon^2 g(\mu,\kappa)- \tau^2 h(\kappa).
	\end{aligned}
\end{equation}
The above system governs the evolution of the mean orientation $\mu(t)$ and of the \color{black} order parameter \color{black} \color{black}$\kappa(t)$, which parametrize the approximating probability distribution $\tilde p(\theta|\mu(t),\kappa(t))$. We recall that the functions $f$, $g$, and $h$ have been defined in \eqref{eq:def-fgh}. We also recall that, as stated in the Introduction, we restrict attention to the case of a single-well energy, consistent with our choice of a unimodal approximate probability distribution. For definiteness, throughout this discussion, we shall restrict attention to the parameter regime
\begin{equation}\label{eq:choice}
	k\ge 1, \qquad \bar k = 1,
\end{equation}
so that \eqref{U} becomes
	\begin{equation}\label{U2}
		\mathscr{U}(\theta) = \frac {1} 2 \left(\cos 2\theta + k\right)^2 + c.
	\end{equation}
We first note that System \eqref{ODE} is autonomous, since the oscillatory behaviour of the strain has been averaged out. As a consequence, the phase plane defined by the $(\mu, \kappa)$ variables is foliated by solution trajectories that do not intersect. Each trajectory corresponds to the forward and backward evolution of the variables $(\mu(t),\kappa(t))$ from a given point. To gain insight into the structure of these trajectories, in Fig.~\ref{fig:phaseportrait} we show a representative collection of curves emanating from the horizontal line $\kappa=5$, together with a curve that originates from a point having coordinates $(\mu,\kappa)=(\pi/2,\delta)$ with $\delta>0$ arbitrarily small. 
\paragraph{Qualitative properties of the solutions.} An inspection of the phase diagrams, together with the more detailed analysis in the Appendix shows that the phase diagram has the following properties.\color{black}
\begin{itemize}
	\item Solution curves are symmetric with respect to the axis $\mu=\pi/2$. This is an immediate  consequence of the following properties of $f$ and $g$:
\begin{equation}
	f(\mu,\kappa)=-f(\pi-\mu,\kappa),\qquad g(\mu,\kappa)=g(\pi-\mu,\kappa),
\end{equation}
which imply that the vector field associated to the dynamical system \eqref{ODE-bis} is mirror-symmetric with respect to the vertical axis $\mu=\pi/2$. 
\end{itemize}
\paragraph{Evolution of the mean orientation $\mu$ towards the minimum $\pi/2$ of the energy.}\color{black}
\begin{itemize}
	\item The set $\dot\mu=0$ consists of the three-vertical lines $\mu=0$, $\mu=\pi/2$, $\mu=\pi$; furthermore, if $0<\mu<\pi/2$ then $\dot\mu>0$, whereas if $\pi/2<\mu<\pi$ then $\dot\mu<0$; moreover $\mu(t)$ converges asymptotically to $\pi/2$. 
	\item The following asymptotic estimates hold:
	\begin{equation}\label{fgh2}
		\begin{aligned}
			&f(\mu,\kappa)\simeq \begin{cases}
				4k_1 k_2 \frac{\sin 2 \mu}\kappa,&\text{as }\kappa\to  0,\\
				-\mathscr U'(\mu),&\text{as }\kappa\to +\infty.\\
			\end{cases}
		\end{aligned}
	\end{equation}
\end{itemize}
We see from \eqref{fgh2} that when the order parameter $\kappa$ is very large the mean orientation $\mu$ obeys the same ODE \eqref{eq:deterministic} that governs $\theta$ in the deterministic case. On the other hand, for $\kappa$ small the dynamics of $\mu$ becomes degenerate, in the sense that $\mu$ sweeps very quickly towards the energy minumum when the order parameter is very small.
\color{black}
\paragraph{Evolution of the order parameter $\kappa$ towards its stationary value $\kappa_*$.}\color{black}
\begin{itemize}
    \item The order parameter converges asymptotically to the value $\kappa_*=\kappa_*(\tau/\varepsilon,k)$ of the fixed point is the unique solution of the equation
  \begin{equation}\label{eq:kappastar2}
k -\left(\frac \tau\varepsilon\right)^2 \frac{\kappa_*}2-\frac{I_2\left(\frac{\kappa_*}{2}\right)}{I_1\left(\frac{\kappa_*}{2}\right)}=0.
\end{equation}
\item The vertical coordinate $\kappa_*$  of the fixed point is a decreasing function of $\tau/\varepsilon$ and an increasing function of the constant  $k$ which enters the expression \eqref{U2} of the renormalized energy $\mathscr U(\theta)$. Moreover,
the following asymptotic estimates hold:
\begin{equation}\label{eq:kappastar3}
	\begin{aligned}
		&{\kappa_*}\simeq 
		\begin{cases}
			\displaystyle {\sqrt{6}}\frac{\varepsilon}{\tau}&\text{ if }k=1,
			\\
			\displaystyle 2(k-1)\frac{\varepsilon^2 }{\tau^2}&\text{ if }k>1,
		\end{cases}&&\qquad\text{for }  \frac\tau\varepsilon\to 0,\\
		&{\kappa_*}(\rho,k)\simeq {2k}\frac{\tau^2}{\varepsilon^2},&&\qquad\text{for } \frac\tau\varepsilon\to \infty.
	\end{aligned}
\end{equation}
\item The stationary point is at the intersection between the set $\dot\mu=0$ and the set  $\dot\kappa=0$. The latter \color{black} consists of two curves that emanate from the points of coordinates $(\mu,\kappa)=(\frac \pi 4,0)$ and $(\mu,\kappa)=(\frac 3 4\pi,0)$ on  the horizontal axis $\kappa=0$. If $\tau>0$, these curves are bounded, and join at a  unique point $(\mu,\kappa)=(\pi/2,\kappa_*)$ on the middle vertical axis; the resulting curve divides the phase plane into an ``inner region'' (below this curve) where $\dot\kappa>0$, and an ``outer region'' where $\dot\kappa<0$.

\end{itemize}
Fig.~\ref{fig:stationary_state_kappa} shows the plots of the root $\kappa_*(\tau/\varepsilon,k)$ of \eqref{eq:kappastar2} for different values of $k$. These plots confirm the monotone dependence of $\kappa_*$ on the ratio $\tau/\varepsilon$. We note that for $k>1$ the slope of the log-log plot is constant For $\tau / \varepsilon \gtrsim 1$ the three curves collapse almost onto one another, showing that in the diffusion-dominated regime the fixed-point concentration becomes essentially independent of the energetic parameter $k$.

\paragraph{Two-stage reorientation.}
Direct inspection of the phase plane shows that, while the mean concentration $\mu(t)$ evolves monotonically towards its equilibrium value $\pi/2$, the order parameter $\kappa(t)$ may be a non-monotone function of $t$, depending on the initial condition.

\begin{figure}[H]
	\centering
	\includegraphics[width=0.5\linewidth]{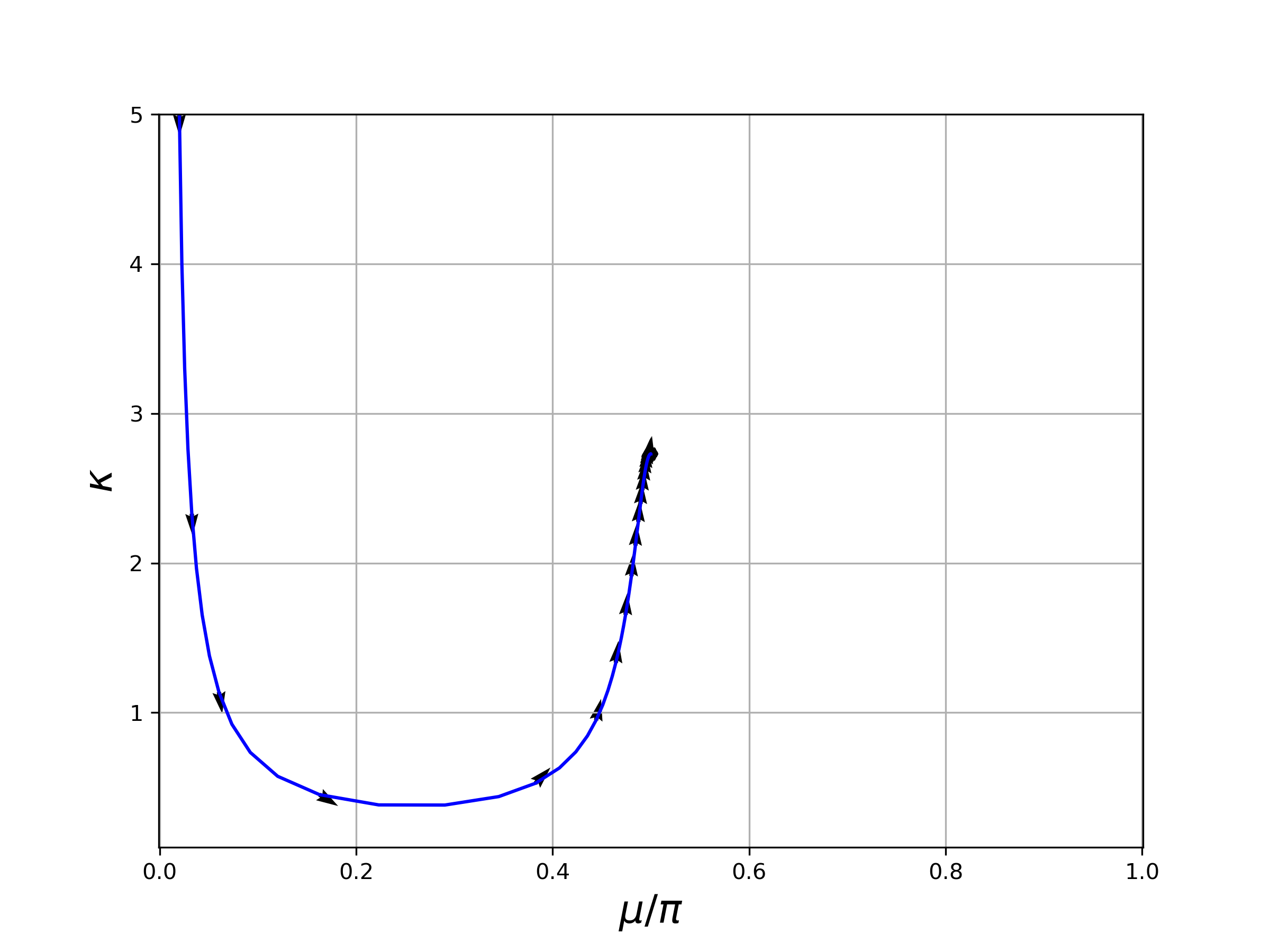}
	\caption{Example of two-stage reorientation. Parameters  are $\varepsilon/\tau=1$, $k=1$. The initial state is $\mu_0/\pi=2\cdot 10^{-2}$, $\kappa_0=5$.}
	\label{fig:phaseportrait-two-stage}
\end{figure}

More generally, if the initial point $(\mu_0,\kappa_0)$ is close to one of the lateral vertical axes $\mu=0$ (or $\mu=\pi/2$), and if the initial concentration $\kappa_0$ is sufficiently large, then the trajectory described by the pair $(\mu(t),\kappa(t))$ on the phase plane starts by going downwards, running almost parallel to the vertical axis $\mu=0$ (or $\mu=\pi$); then, it bends towards the middle vertical axis; eventually it goes upwards towards the stationary state. Thus, if the cell population starts with a narrowly peaked orientation near an energy maximum, the cells do not immediately realign toward the nearest energy minimum. Instead, reorientation takes place in \emph{two stages}: first, the distribution broadens (so that the order parameter \(\kappa(t)\) decreases) and \(\mu(t)\) rapidly converges to the energy minimum at \(\pi/2\); then, the order parameter \(\kappa(t)\) grows again as the distribution refocuses around that minimum. This behaviour is illustrated in Fig.~\ref{fig:phaseportrait-two-stage} as well as Fig.~\ref{fig:kappa40} in Section~\ref{sec:experiments}.

Note that, without the periodic domain in \(\mu\), the points \(\mu=0\) and \(\mu=\pi/2\) would act as  barriers, and this two-stage reorientation would not occur.

\begin{remark}[Proposal for a new experiment to test the model]
	A possible experiment to test the model may consist in the following two steps. First, apply a cyclic stretch along the $x$-axis until the cell population reaches a steady-state alignment along the $y$-axis. Then, switch the stretching direction to the $y$-axis and continue until cells realign along the $x$-axis. According to our predictions, immediately after changing the stretch axis, the angular distribution should briefly broaden, before refocusing into a tight distribution around the new minimal-energy direction. 
\end{remark} \color{black}

\begin{remark}[Solvability of \eqref{eq:kappastar2}]
    To see that \eqref{eq:kappastar2} has unique solution, is suffices to notice that the ratio ${I_2\left(z\right)}/{I_1\left(z\right)}$ is an increasing continuous function which vanishes for $z=0$ and tends to $1$ for $z\to\infty$. Thus, the left-hand side of \eqref{eq:kappastar2} is a decreasing function of $\kappa_*$.  This function is equal to $k$ for $\kappa_*=0$ is negative for $\kappa_*$ sufficiently large. Thus, it has a unique zero. 
    The monotonicity properties of $\kappa_*$ follow upon observing that the left-hand side of \eqref{eq:kappastar2} is a decreasing function of $\tau/\varepsilon$ and an increasing function of $k$. A more detailed discussion is in the Appendix.
\end{remark}
\color{black}

\section{Numerical tests}\label{sec:test}
In Section \eqref{sec:model} we introduced the approximate probability distribution $\tilde {p}(\theta;\mu(t),\kappa(t))$ defined in \eqref{p-tilda}, where the functions $\mu(t)$ and $\kappa(t)$ (resp. mean orientation and order parameter at time $t$)   are the solution of the ODE system \eqref{ODE} with the initial conditions \eqref{minimization}. In this section we assess the quality of the approximation $p(\theta,t)\simeq \tilde p(\theta|\mu(t),\kappa(t))$ through some numerical tests.

\paragraph{Stationary solutions.}
Our first comparison concerns the stationary solution $p_\infty(\theta)$ (\emph{cf.} \eqref{equilibrium}) of the Fokker--Planck equation \eqref{fp} and the approximation $\tilde p(\theta|\pi/2,\kappa_*)$, which corresponds to the stationary state of the dynamical system \eqref{ODE}. We define $\overline\kappa_*$ by solving the minimization problem
\begin{equation}\label{minimization3}
\overline\kappa_*=\operatorname{argmin}_{\overline\kappa}D_{\mathrm{KL}}(p_\infty\| \tilde{p}(\,\cdot\,| \pi/2, \overline\kappa)),
\end{equation}
where we recall that $D_{\rm KL}$ is the Kullbach--Leibler divergence introduced in \eqref{def:DKL}. The probability distribution $\tilde{p}(\theta| \pi/2, \overline\kappa_*)$ is the best information-theoretic approximation of $p_\infty(\theta)$ in the family $\mathscr A$ of approximating probability distributions defined in \eqref{eq:family}. We note that  $p_\infty(\theta)$ depends on $\tau/\varepsilon$ and on $k$ (through $\mathscr U(\theta)$). Thus, $\overline\kappa_*$ depends on the parameters $\tau/\varepsilon$ and $k$. These are the same parameters that defermine $\kappa_*$. 

\begin{figure}[H]
    \centering
        \includegraphics[width=0.5\textwidth]{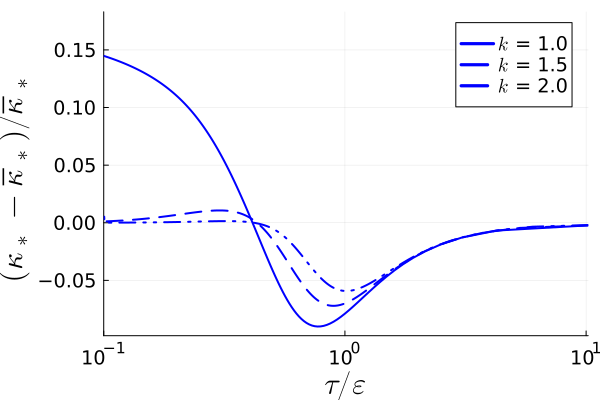}
        \caption{Relative error between $\kappa_*$ and $\overline{\kappa}_*$.}
        \label{fig:relative_error}
\end{figure}

\begin{figure}[H]
    \centering
    \begin{subfigure}[b]{0.49\textwidth}
    \centering
	\includegraphics[width=1\linewidth]{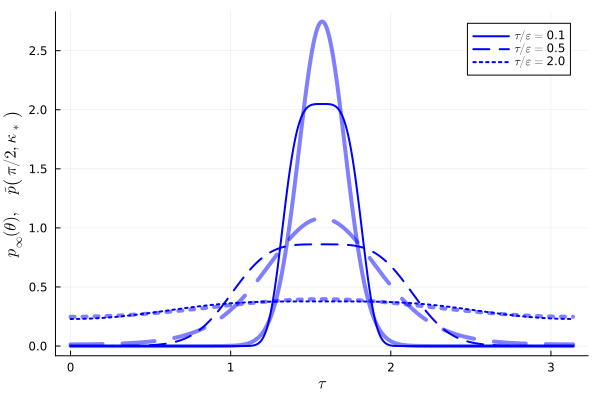}
    \caption{}
    \end{subfigure}
     \begin{subfigure}[b]{0.49\textwidth}
     \centering
    \includegraphics[width=1\linewidth]{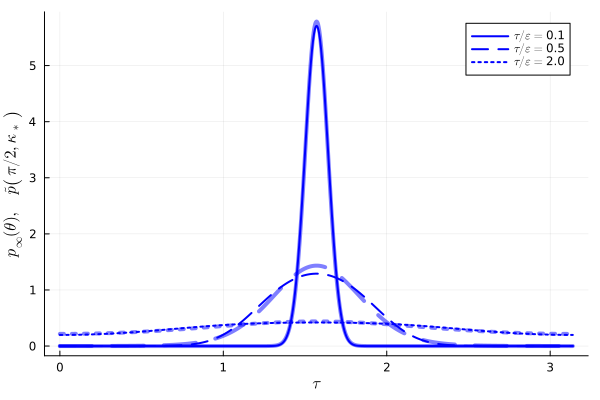}
    \caption{}
    \end{subfigure}
    \caption{Equilibrium distributions: von Mises approximation (thin) vs. exact (thick transparent) for $k=1$ (a) and $k=1.5$ (b).}
    \label{fig:distribution_comparison}
\end{figure}

Fig.~\ref{fig:relative_error} shows the relative error between $\kappa_*$ and $\overline\kappa_*$. The agreement improves as $k$ increases, and the smallest relative error is for $\tau/\varepsilon$ close to 1.
Fig.~\ref{fig:distribution_comparison} shows a comparison between $p_\infty(\theta)$ and $\widetilde p(\theta|\pi/2,\kappa_*)$ for different values of the ration $\tau/\varepsilon$ when $k=1$. Although the error may be not negligible if one uses the sup norm, the quantities of interest, namely the true order parameters $\overline\kappa_*$ and its approximation $\kappa_*$, are relatively close to each other, especially for $k>1$.
\paragraph{Transient regime.} Next, we compare transient solutions. To this aim, we employ a standard numerical scheme --- specifically, the method of lines in our implementation --- to construct a numerical solution of the Fokker--Planck equation \eqref{fp}. Then, at each time $t$, we define $\overline\mu(t)$ and $\overline\kappa(t)$ by solving the minimization problem\color{black}
\begin{equation}\label{minimization2}
\big(\overline\mu(t),\overline\kappa(t)\big)=\operatorname{argmin}_{\big(\overline\mu,\overline\kappa\big)}D_{\mathrm{KL}}(p(\cdot, t) \| \tilde{p}(\cdot ; \overline\mu, \overline\kappa)),
\end{equation}
using the same procedure adopted in Section~\ref{sec:initial_conditions} to infer the initial values $\mu_0$ and $\kappa_0$ from the initial probability distribution $p_0(\theta,t)$. 

The probability distribution $\tilde{p}(\theta ; \overline\mu(t), \overline\kappa(t))$ represents the best information-theoretic approximation of $\tilde p(\theta,t)$  in the  family $\mathscr A$ of approximating distributions defined in \eqref{eq:family}. Accordingly, the functions $\overline\mu(t)$ and $\overline\kappa(t)$ are the optimal information-theoretic estimates of the mean orientation and of the order parameter at time $t$. Their agreement with $\mu(t)$ and $\kappa(t)$ serves as our metric for assessing the quality of our approximation.
\color{black}

\begin{figure}[H]
\centering
\begin{subfigure}[b]{0.45\textwidth}
	\includegraphics[width=\textwidth]{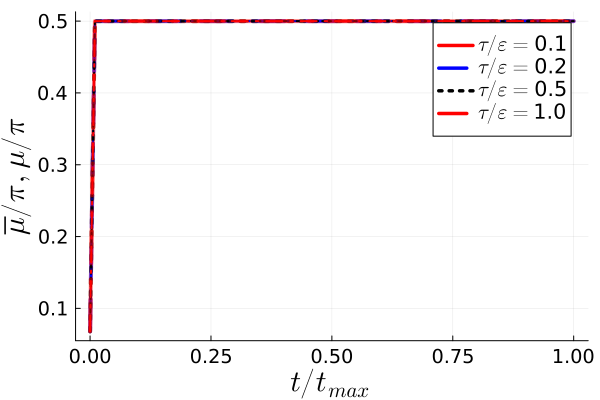}
	\caption{}
\end{subfigure}	
\qquad
\begin{subfigure}[b]{0.45\textwidth}
	\includegraphics[width=\textwidth]{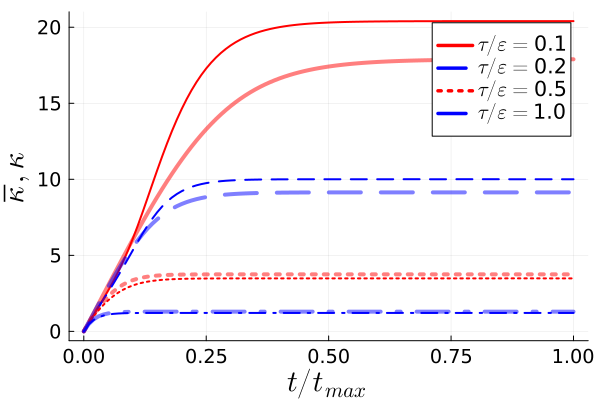}
	\caption{}
	\end{subfigure}
\caption{Comparison between the mean orientations (a) and the order parameters (b) extrapolated from the solution of the Fokker--Planck equation \eqref{fp} (thick, transparent lines) and the solution of the ODE system \eqref{ODE} (thin lines) for uniform initial condition. The plots of $\mu$ are all superposed and instantaneously converge towards the equilibrium value $\pi/2$.}
\label{fig:mu_kappa}
\end{figure}

\begin{figure}[H]\label{fig:tildas4}
\centering
\begin{subfigure}[b]{0.45\textwidth}
	\includegraphics[width=\textwidth]{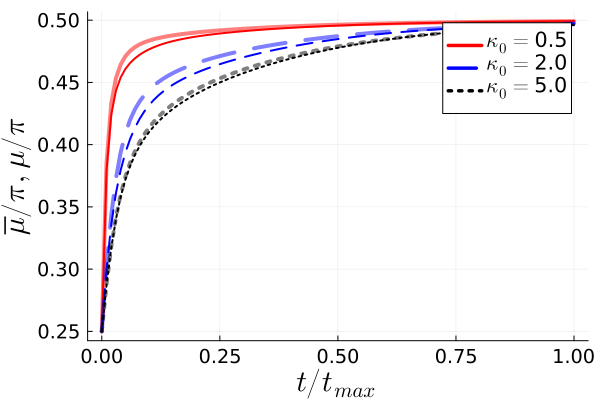}
	\caption{}
	\label{fig:mu4}
\end{subfigure}
\hfill
\begin{subfigure}[b]{0.45\textwidth}
	\includegraphics[width=\textwidth]{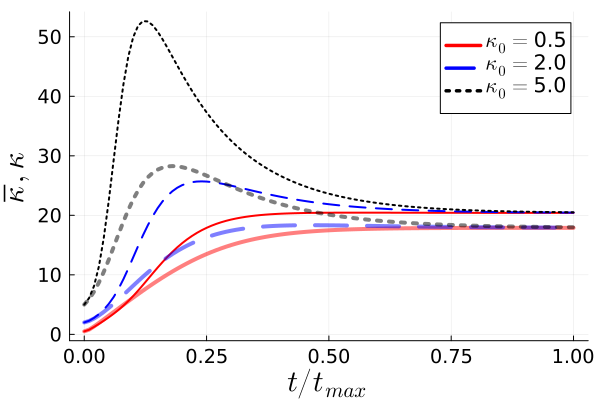}
	\caption{}
	\label{fig:kappa4}
\end{subfigure}
\caption{Same as Fig.~\ref{fig:mu_kappa}, but with initial conditions $\mu_0=\pi/4$ and $\kappa_0=0.5, 2.0, 5.0$.}
\label{fig:mu_kappa4}
\end{figure}

The following parameters have been used in our simulation:
\begin{itemize}
\item Main parameters: $\varepsilon=1$, $K=1$, $\tau=0.1$, and $\eta=1$.
\item Additional model parameters:  $K_s=0.7$,  $K_{\perp} = \frac{K_s - 1 + \alpha_L}{1 + \alpha_L}=0.28$, $r = \frac{1 - \alpha_L}{1 + \alpha_L}=0.11$ with $\alpha_L = 0.8$ (see \cite{loy2023statistical}).
\end{itemize}
The above choice of parameters yields $k=1.0$.
Fig~\ref{fig:mu_kappa} shows the plots of $\mu(t)$, $\bar\mu(t)$, $\kappa(t)$ and $\bar\kappa(t)$ when the initial probability distribution is uniform, for different choices of the ratio $\tau/\varepsilon$. Consistent with Fig.~3, higher values of $\tau/\varepsilon$ correspond to lower values of the equilibrium order parameter. We recall that when the probability distribution is uniform the mean orientation $\mu_0$ is not defined uniquely. Here we have chosen $\mu_0=0$. 
Fig.~\ref{fig:mu_kappa4} show the results of our simulations when the initial probability distribution is $p_0(\theta)=\tilde p(\theta|\mu_0,\kappa_0)$ with $\mu_0=\pi/4$ (the boundary of the spinodal region), for different values of $\kappa_0$.  

\begin{figure}[H]
\centering
\begin{subfigure}[b]{0.45\textwidth}
	\includegraphics[width=\textwidth]{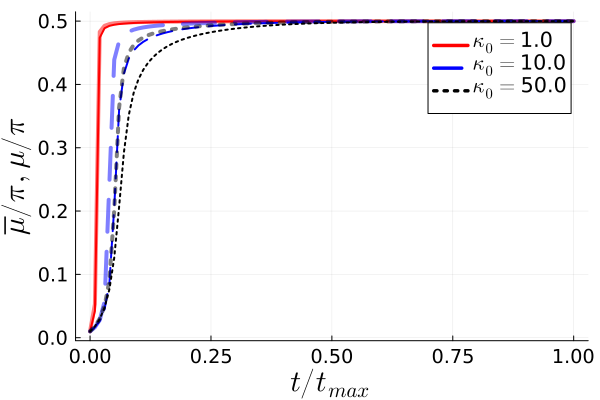}
	\caption{}
	\label{fig:mu4_0}
\end{subfigure}
\hfill
\begin{subfigure}[b]{0.45\textwidth}
	\includegraphics[width=\textwidth]{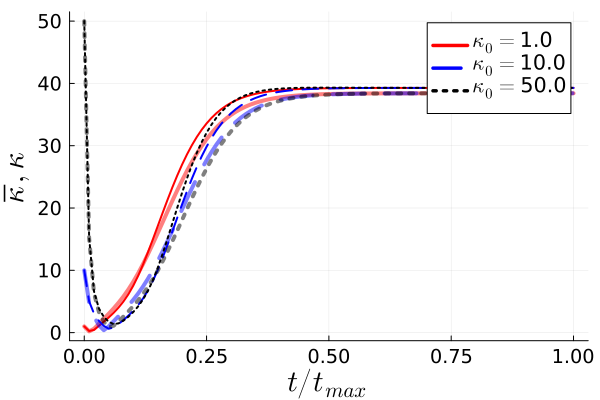}
	\caption{}
	\label{fig:kappa40}
\end{subfigure}
\caption{Same as Fig.~\ref{fig:mu_kappa}, but with initial conditions $\mu_0=\pi/100$ and $\kappa_0=1, 10, 50$.}
\label{fig:mu_kappa0}
\end{figure}

\color{black}Finally, Fig.~\ref{fig:mu_kappa0} shows the evolution of $\mu(t)$ and $\kappa(t)$ when $\mu_0$ is close to the energy maximum, with different values of the initial concentration. 	In connection with the discussion at the end of Section~\ref{sec:qualitative}, we notice from Fig.~\ref{fig:kappa40} that two-stage orientation is also a phenomenon of the original model based on the full Fokker--Planck equation.\color{black}

Both in Fig~\ref{fig:mu_kappa} and Fig.~\ref{fig:mu_kappa4} the mean orientation quickly settles on the minimum of the energy, at $\pi/2$. However, the concentration displays two remarkably different behaviors. In the first case, it increases towards a stationary value. In the second case, for $\kappa=5.0$ it first increases, and then decreases, settling quickly towards its asymptotic limit. \color{black}This behaviour is further discussed in Remark~\ref{rem:nonmonotone} in the Appendix.\color{black}

 In all simulations, with a couple of exceptions, the plots of $\mu(t)$ and $\kappa(t)$ match quite closely those of their information-theoretical counterparts $\overline\mu(t)$ and $\overline\kappa(t)$. These results suggest that both the mean orientation and the order parameter computed using the reduced model \eqref{ODE} are a good approximation of the values that minimize the K--L divergence.

\begin{remark}[Non-monotone behaviour of $\kappa$]
	\color{black}In Fig.~\ref{fig:kappa4} \color{black} the order parameter first grows and then decreases for $\kappa_0=5.0$. The reason for this behaviour is that the set $\dot\kappa>0$, depending on the parameters, may have two pronounced lobes (see Fig.~\ref{fig:phaseportrait3} in the Appendix). If the initial state is in one of these lobes, then the concentration increases until the trajectory in phase space intersects the set $\dot\kappa=0$. After that point, the order parameter decreases.
\end{remark}
\begin{remark}[Interpretation of our numerical test]
In exact terms, our test does not aim to quantify the error between the true solution $p(\theta,t)$ and the approximation $\tilde p(\theta|\mu(t),\kappa(t))$. Instead, it seeks to demonstrate that the solution $(\mu(t),\kappa(t))$ of the dynamical system \eqref{ODE} closely matches the optimal parameters $\overline\mu(t)$ and $\overline\kappa(t)$ obtained via an information-theoretic criterion.
\end{remark}
\color{black}

\section{Comparison with experimental data}\label{sec:experiments}
\begin{figure}[H]
	\centering
	\includegraphics[width=0.5\linewidth]{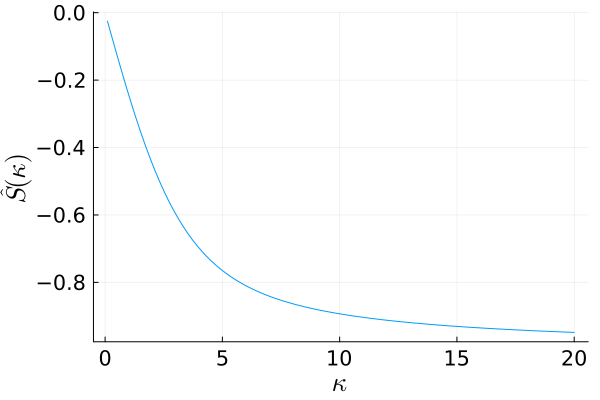}
	\caption{Plot of the function $\hat S(\kappa)=-{I_1\left(\frac{\kappa}{2}\right)}/{I_0\left(\frac{\kappa}{2}\right)}$.}
	\label{fig:plothatS}
\end{figure}

\color{black}Mao et al.~\cite{maoCriticalFrequencyCritical2021} experimentally investigated the reorientation dynamics of human mesenchymal stem cells subjected to cyclic stretching using a microfluidic device. They quantified the degree of ordering through the expected value
\begin{equation}\label{eq:order}
	S(t) = \langle \cos 2\theta \rangle_{p_{\rm emp}(\cdot,t)},
\end{equation}
of $\cos\theta$ with respect to \emph{empirical probability distribution} $p_{\rm emp}(\theta,t)$, that is, the distribution that best fits the available experimental data. In our model, the expected value of $\cos\theta$ is
\begin{equation}\label{Smao}
	\tilde S(t)= \langle \cos 2\theta \rangle_{\tilde p(\cdot|\mu(t),\kappa(t))},
\end{equation}
where  $\mu(t)$ and $\kappa(t)$ are the solutions of \eqref{ODE}, with initial condition corresponding to the uniform probability distribution, as discussed in Section~\ref{sec:initial_conditions}, Remark~\ref{rem:specialCases}. As we have seen in Section~\ref{sec:experiments}, with such initial conditions the average orientation $\mu(t)$ quickly settles to $\pi/2$, with a dynamics that is much faster than that of $\kappa(t)$. Accordingly, 
\begin{equation}\label{eq:approx4}
	\tilde S(t)\simeq \hat S(\kappa(t)),
	\end{equation}
where $\hat S(\kappa)=\langle \cos 2\theta \rangle_{\tilde p(\cdot|\pi/2,\kappa)}$ is the expectation of $\cos\theta$ with respect to the probability density $\tilde p(\theta|\pi/2,\kappa)$. A direct calculation leads to 
\begin{equation}	
\hat S(\kappa)=-\frac{I_1\left(\frac{\kappa}{2}\right)}{I_0\left(\frac{\kappa}{2}\right)},
\end{equation}
where we recall that $I_0$ and $I_1$ are modified Bessel functions of the first kind. The function $\hat S(\kappa)$ is monotone decreasing, with $\hat S(0)=0$, and $\hat S(\kappa)\to -1$ as $\kappa\to\infty$, confirming the monotone correlation between our order parameter $\kappa$ and the order parameter $S$ adopted in \cite{maoCriticalFrequencyCritical2021}. An illustrative plot of $\hat S(\kappa)$ is shown in Fig.~\ref{fig:plothatS}. 
The experimental data regarding the evolution of $S(t)$ is reported in \cite[Fig.~5]{maoCriticalFrequencyCritical2021}, assuming that the \emph{strain amplitude} $\varepsilon_{\rm max}$, defined by
\begin{equation}
	\varepsilon_{\rm max}=2\varepsilon_{\rm avg}=2\sqrt{2/3}\,\varepsilon,
\end{equation}
attains the values $0.02$, $0.05$, and $0.1$. We used these data to tune the three parameters of our model, namely, the dimensionless parameters $\tau$ and $k$ (see \eqref{diffusion} and \eqref{U}) which govern steady states, and the characteristic time $K/\eta$ (see \eqref{Omega}), which affects the relaxation dynamics, so that  $\tilde S(t)$ would match the experimental data on $S(t)$.

The parameter $\tilde S(t)$ starts from $0$, is monotone decreasing, and its stationary value is $\hat S(\kappa_*)$, where $\kappa_*$ depends on $\tau/\varepsilon$ and on $k$, by \eqref{eq:kappastar2}. 
By taking
\begin{equation}
\tau=0.04,\qquad  k=2,	
\end{equation}
we managed to obtain a good matching between $\hat S(\kappa_*)$ and the stationary values of $S(t)$ observed from experiment.

\begin{figure}[H]
	\centering
	\begin{subfigure}[b]{0.4\textwidth}
		\includegraphics[width=\textwidth]{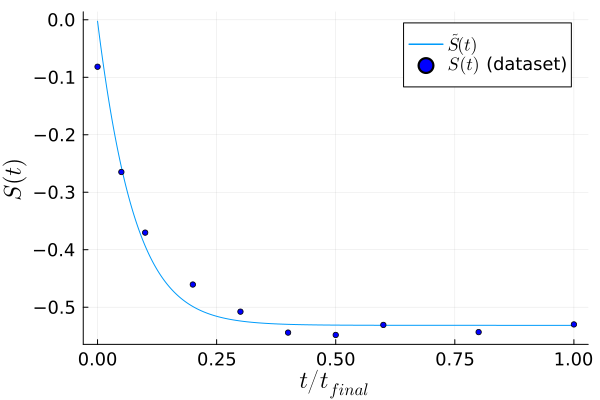}
		\caption{$\varepsilon_{\text{max}} = 0.02$ and $\eta/K = 2.0$.}
		\label{fig:evolutionsourmodel002}
	\end{subfigure}
\qquad 
	\begin{subfigure}[b]{0.4\textwidth}
		\includegraphics[width=\textwidth]{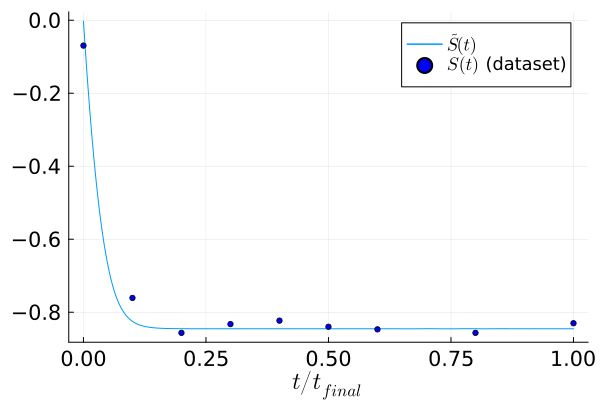}
		\caption{$\varepsilon_{\text{max}} = 0.05$ and $\eta/K = 3.5$.}
		\label{fig:evolutionsourmodel005}
	\end{subfigure}
	\\
	\begin{subfigure}[b]{0.4\textwidth}
		\includegraphics[width=\textwidth]{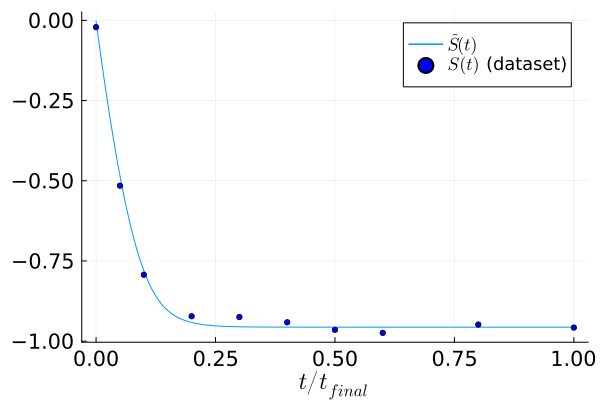}
		\caption{$\varepsilon_{\text{max}} = 0.1$ and $\eta/K = 30$.}
		\label{fig:evolutionsourmodel01}
	\end{subfigure}
	\caption{Time evolution of the order parameter $\tilde{S}(\kappa(t))$, as predicted by \eqref{ODE} for different values of $\varepsilon_{\text{max}}$ and $\eta/K$.}
	\label{fig:evolution_combined}
\end{figure}

Next, we tuned the characteristic time $\eta / K$ so that the transient behaviour of $\tilde S(t)$ would replicate that of $S(t)$. We found that a good match could be obtained by assuming that the characteristic time $\eta/K$ depends on the strain amplitude. Figure \ref{fig:evolution_combined} shows the comparison between the dataset from \cite[Fig.~5]{maoCriticalFrequencyCritical2021} and $\tilde S(t)$ with $\eta/K=2.0$ for $\varepsilon_{\rm max}=0.02$, $\eta/K=3.5$ for $\varepsilon_{\rm max}=0.05$, and $\eta/K=30$ for $\varepsilon_{\rm max}=0.1$. This finding suggests that the viscosity constant should change with the amplitude, potentially indicating the onset of non-linear effects.
\color{black}

\section{Conclusions}\label{sec:conclusions}
\color{black}
We have shown that the full Fokker-Planck description of cell-orientation under cyclic stretch can be approximated by a two-dimensional autonomous dynamical system for the mean orientation $\mu(t)$ and an order parameter $\omega(t)$. This reduced model captures both the transient reorientation kinetics and the steady-state orientation distributions, while being amenable to phase-plane analysis.\color{black}
\medskip

The insight provided by phase-plane analysis uncovers an interesting phenomenon, which we call \emph{two-stage reorientation}: when cells initially share an orientation corresponding to an energy maximum, their evolution towards the energy minimum is not achieved through a collective, ordered reorientation; instead their collective direction first spreads out into a broad distribution, and then concentrates again at the energy minimum. \color{black}This observation suggests a new experiment to test the theoretical framework on which the present paper is based.\color{black}

\medskip

Comparison with experimental data on transient dynamics suggests to us that taking the viscosity $\eta$ as a constant is too crude an assumption. In particular, we think that $\eta$ should be an increasing function of the strain amplitude. This does not mean that cells remodel their cytoskeleton more slowly when subjected to larger cyclic deformations but that assuming a constant viscosity $\eta$ would imply a too-fast relaxation dynamics.
\color{black}
\medskip

Our numerical results indicate that the scheme proposed in \cite{reina2023statistical} implicitly minimizes the Kullback--Leibler divergence between the exact solution and its approximation. This observation find its confirmation in the recent contribution \cite{leadbetterStructurePreservingClosure2025}.

\medskip
It has been observed in \cite{maoCriticalFrequencyCritical2021} that 
the degree of spreading increases if the frequency of the cyclic strain decreases. Thus the spreading of the orientation angle is reduced not only when the strain amplitude increases, but also when the frequency of oscillation increases. Our model could be made consistent with this observation by incorporating viscoelastic effects, as in \cite{lucci2021stability} (see also the discussion in \cite{loy2023statistical}).
\medskip 

Our analysis may be extended to the case when the energy features two minima, as occurs when the biaxiality ration $r$ is sufficiently close to 1. In this case, the most natural modification of our ansatz would be to consider a bimodal von Mises distribution, which displays two peaks \cite{gattoGeneralizedMisesDistribution2007}. Alternatively, one could consider a superposition of two von Mises distributions. While symmetry assumptions, leveraging the energy's mirror-symmetric nature, might reduce this to a two-dimensional system, in both instances we have not been able to compute the coefficients of the governing ODE system in terms of known functions.

\section{Acknowledgements}
GT thanks Giulio Lucci for suggesting the use of the von Mises distribution. GT also thanks Riccardo Durastanti and Lorenzo Giacomelli for useful discussions on the subject of this manuscript. GT acknowledges support from the Italian Ministry of University and Research through project PRIN 2022NNTZNM DISCOVER, and the “Departments of Excellence” initiative. The hardware used for the numerical calculations was acquired through support from the Rome Technopole Foundation. GS thanks Emilio Cirillo for useful discussions on the subject of stochastic systems.  
GS is supported by the European Union - Next Generation EU PRIN 2022 research project ''The Mathematics and Mechanics of nonlinear wave propagation in solids'' (grant n.2022P5R22 A). RA thanks MISTI -- Italy for their support. Support from INdAM-GNFM is also acknowledged.

\bibliographystyle{abbrv}
\bibliography{biblio.bib}

\begin{thebibliography}{10}

\bibitem{abeyaratneElementaryModelFocal2022}
R.~Abeyaratne, E.~Puntel, and G.~Tomassetti.
\newblock An {{Elementary Model}} of {{Focal Adhesion Detachment}} and
  {{Reattachment During Cell Reorientation Using Ideas}} from the {{Kinetics}}
  of {{Wiggly Energies}}.
\newblock {\em J. Elasticity}, 155, 2022.

\bibitem{abramowitz1964handbook}
M.~Abramowitz and I.~A. Stegun, editors.
\newblock {\em Handbook of Mathematical Functions with Formulas, Graphs, and
  Mathematical Tables}, volume~55 of {\em Applied Mathematics Series}.
\newblock National Bureau of Standards, Washington, D.C., 1964.

\bibitem{barronEffectPhysiologicalCyclic2007}
V.~Barron, C.~Brougham, K.~Coghlan, E.~McLucas, D.~O'Mahoney, C.~Stenson-Cox,
  and P.~E. McHugh.
\newblock The effect of physiological cyclic stretch on the cell morphology,
  cell orientation and protein expression of endothelial cells.
\newblock {\em J. Mater. Sci. Mater. Med.}, 18:1973--1981, 2007.

\bibitem{buck1979longitudinal}
R.~C. Buck.
\newblock The longitudinal orientation of structures in the subendothelial
  space of rat aorta.
\newblock {\em American Journal of Anatomy}, 156(1):1--14, 1979.

\bibitem{buck1980reorientation}
R.~C. Buck.
\newblock Reorientation response of cells to repeated stretch and recoil of the
  substratum.
\newblock {\em Experimental Cell Research}, 127(2):470--474, 1980.

\bibitem{collinsworthOrientationLengthMammalian2000}
A.~M. Collinsworth, C.~E. Torgan, S.~N. Nagda, R.~J. Rajalingam, W.~E. Kraus,
  and G.~A. Truskey.
\newblock Orientation and length of mammalian skeletal myocytes in response to
  a unidirectional stretch.
\newblock {\em Cell Tissue Res.}, 302:243--251, 2000.

\bibitem{dasCellReorientationCyclically2022}
S.~Das, A.~Ippolito, P.~McGarry, and V.~S. Deshpande.
\newblock Cell reorientation on a cyclically strained substrate.
\newblock {\em PNAS Nexus}, 1:pgac199, Nov. 2022.

\bibitem{de2007dynamics}
R.~De, A.~Zemel, and S.~A. Safran.
\newblock Dynamics of cell orientation.
\newblock {\em Nature Physics}, 3:655--659, 2007.

\bibitem{eastwoodEffectPreciseMechanical1998}
M.~Eastwood, V.~C. Mudera, D.~A. McGrouther, and R.~A. Brown.
\newblock Effect of precise mechanical loading on fibroblast-populated collagen
  lattices: Morphological changes.
\newblock {\em Cell Motil. Cytoskeleton}, 40:13--21, 1998.

\bibitem{faustCyclicStressMHz2011}
U.~Faust, N.~Hampe, W.~Rubner, N.~Kirchge{\ss}ner, S.~Safran, B.~Hoffmann, and
  R.~Merkel.
\newblock Cyclic {{Stress}} at {{mHz Frequencies Aligns Fibroblasts}} in
  {{Direction}} of {{Zero Strain}}.
\newblock {\em PLoS ONE}, 6:e28963, 2011.

\bibitem{gattoGeneralizedMisesDistribution2007}
R.~Gatto and S.~R. Jammalamadaka.
\newblock The generalized von {{Mises}} distribution.
\newblock {\em Stat. Method.}, 4:341--353, July 2007.

\bibitem{jufri2015mechanical}
N.~F. Jufri, A.~Mohamedali, A.~Avolio, and M.~S. Baker.
\newblock Mechanical stretch: physiological and pathological implications for
  human vascular endothelial cells.
\newblock {\em Vascular Cell}, 7(1):8, 2015.

\bibitem{kullback1959information}
S.~Kullback.
\newblock {\em Information Theory and Statistics}.
\newblock John Wiley \& Sons, New York, 1959.
\newblock Reprinted by Dover Publications, 1997.

\bibitem{kullback1951information}
S.~Kullback and R.~A. Leibler.
\newblock On information and sufficiency.
\newblock {\em Annals of Mathematical Statistics}, 22(1):79--86, 1951.

\bibitem{reina2023statistical}
T.~Leadbetter, P.~K. Purohit, and C.~Reina.
\newblock A statistical mechanics framework for constructing non-equilibrium
  thermodynamic models.
\newblock {\em PNAS Nexus}, 2(2), 2023.

\bibitem{leadbetterStructurePreservingClosure2025}
T.~Leadbetter, P.~K. Purohit, and C.~Reina.
\newblock On a structure preserving closure of {{Langevin}} dynamics,
  arXiv:2506.08156, 2025.

\bibitem{livne2014cell}
A.~Livne, E.~Bouchbinder, and B.~Geiger.
\newblock Cell reorientation under cyclic stretching.
\newblock {\em Nature Communications}, 5:3938, 2014.

\bibitem{loy2023statistical}
N.~Loy and L.~Preziosi.
\newblock A statistical mechanics approach to describe cell reorientation under
  stretch.
\newblock {\em Bulletin of Mathematical Biology}, 85(60), 2023.

\bibitem{lucci2021stability}
G.~Lucci, C.~Giverso, and L.~Preziosi.
\newblock Cell orientation under stretch: Stability of a linear viscoelastic
  model.
\newblock {\em Mathematical Biosciences}, 337:108630, 2021.

\bibitem{lucci2021nonlinear}
G.~Lucci and L.~Preziosi.
\newblock A nonlinear elastic description of cell preferential orientations
  over a stretched substrate.
\newblock {\em Biomechanics and Modeling in Mechanobiology}, 20:631--649, 2021.

\bibitem{maoCriticalFrequencyCritical2021}
T.~Mao, Y.~He, Y.~Gu, Y.~Yang, Y.~Yu, X.~Wang, and J.~Ding.
\newblock Critical frequency and critical stretching rate for reorientation of
  cells on a cyclically stretched polymer in a microfluidic chip.
\newblock {\em ACS Applied Materials \& Interfaces}, 13(12):13934--13948, 2021.

\bibitem{MardiaJupp2000}
K.~V. Mardia and P.~E. Jupp.
\newblock {\em Directional Statistics}.
\newblock Wiley Series in Probability and Statistics. John Wiley \& Sons,
  Chichester, 2000.

\bibitem{riskenFokkerPlanckEquationMethods1996}
H.~Risken.
\newblock {\em The {{Fokker-Planck}} Equation: Methods of Solution and
  Applications}.
\newblock Springer-Verlag, New York, 1996.

\bibitem{shannonMathematicalTheoryCommunication1948}
C.~E. Shannon.
\newblock A {{Mathematical Theory}} of {{Communication}}.
\newblock {\em Bell System Technical Journal}, 27:379--423, July 1948.

\bibitem{shishvan2018homeostatic}
S.~S. Shishvan, A.~Vigliotti, and V.~S. Deshpande.
\newblock The homeostatic ensemble for cells.
\newblock {\em Biomechanics and Modeling in Mechanobiology}, 17:1631--1662,
  2018.

\bibitem{vonMises1918reprint}
R.~von Mises.
\newblock {\"U}ber die ``ganzzahligkeit'' der atomgewichte und verwandte
  fragen. {R}eprinted from \textit{Physikalische Zeitschrift} 19 (1918), pp.
  490--500.
\newblock In P.~Frank, S.~Goldstein, M.~Kac, W.~Prager, G.~Szeg{\"o}, and
  G.~Birkhoff, editors, {\em Selected Papers of Richard von Mises: Volume 2:
  Probability and Statistics, General}, pages 123--133. American Mathematical
  Society, Providence, RI, 1964.
\newblock Reprinted from \textit{Physikalische Zeitschrift} 19 (1918), pp.
  490--500.

\end{thebibliography}

\section*{Appendix.}
In this section we perform a detailed study of the phase plane and we prove the assertions given in Section~\ref{sec:qualitative}. We also discuss additional properties of the phase that are not mentioned in the main body of this paper.
\subsection*{The evolution of $\mu(t)$} We begin by studying the first of \eqref{ODE-bis}. We note that
\begin{equation}
	f(\mu,\kappa)=f(\mu+\pi,\kappa),\qquad g(\mu,\kappa)=g(\mu+\pi,\kappa),
\end{equation}
that is, $f(\mu,\kappa)$ and $g(\mu,\kappa)$ are $\pi$-periodic with respect to $\mu$. Since $f(\mu,\kappa)$ and $g(\mu,\kappa)$ are also mirror-symmetric with respect to $\mu=\pi/2$, we have, as a consequence, 
\begin{equation}\label{eq:fixedmu}
	f(0,\kappa)=f(\pi/2,\kappa)=f(\pi,\kappa)=0,
\end{equation}
which implies that if $\mu(0)\in \{0,\pi/2,\pi\}$ then $\mu(t)$ is constant along the trajectory. In particular, all fixed points, if any, must lie in the three vertical lines at the extremes and at the center of the phase plane. This, again, in accordance with the evidence from Fig.~\ref{fig:phaseportrait}.

\begin{figure}[H]
\centering
\includegraphics[width=0.4\linewidth]{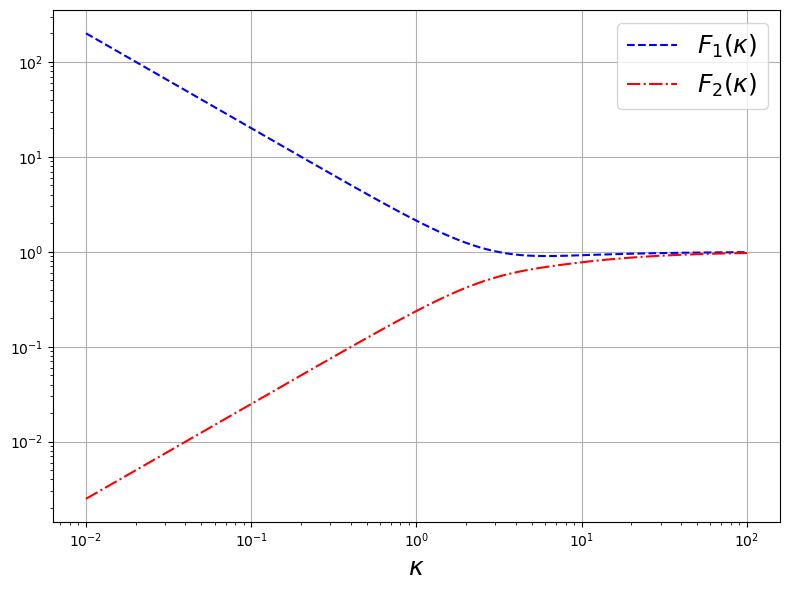}
\caption{Log-log plots of $F_1(\kappa)$ and $F_2(\kappa)$.}\label{fig:plots}
\end{figure}

It is also noted from Fig.~\ref{fig:phaseportrait} that all curves converge towards the axis $\mu=\pi/2$.
To confirm that this is true for every choice of parameters consistent with \eqref{eq:choice}, we need to investigate more in details the properties of $f(\mu,\kappa)$, whose expression, which has been already given in \eqref{eq:def-fgh}, we repeat below:
\begin{equation}
	f(\mu,\kappa) 
	= 2k_1 F_1(\kappa)
	\left(
	F_2(\kappa) \cos 2\mu +
	k_2 \right) \sin 2\mu.	
\end{equation}
The functions $F_1(\kappa)$ and $F_2(\kappa)$, which we plot in Fig.~\ref{fig:plots}, 
are positive. Moreover $F_2(\kappa)$ is increasing with respect to $\kappa$ and smaller than 1. 
\noindent Since we are assuming $k_2\ge 1$, and since $F_2(\kappa)$ is smaller than $1$, we also have that $f(\mu,\kappa)$ has the same sign as $\sin 2\mu$ for all $\kappa$. As a consequence,
\begin{equation}
\begin{aligned}
&f(\mu,\kappa)>0\text{ for }	 0<\mu<\pi/2,\\
&f(\mu,\kappa)<0\text{ for }	 \pi/2<\kappa<\pi.
\end{aligned}
\end{equation}
Therefore, any stable fixed point cannot lie the two extreme vertical lines of the phase plane. 

In order to guarantee that all curves converge to the vertical axis $\pi/2$ in finite time, we need to carefully examine the converging rate of $\mu(t)$, which depends on the factors that multiply $\sin 2\mu$. To this aim, we consider the asymptotic behaviour of $f(\mu,\kappa)$ for small and large $\kappa$. By making use of well-known asymptotic expansions of the modified Bessel functions $I_0(z)$, $I_1(z)$, and $I_2(z)$, which are documented in standard mathematical references (see for instance \cite{abramowitz1964handbook}), we find that the quotients $\frac{I_2(\kappa / 2)}{I_1(\kappa / 2)}$ and $\frac{I_1(\kappa / 2)}{I_0(\kappa / 2)}$ are increasing with $\kappa$, and satisfy
\begin{equation}\label{eq:quotients1}
\frac{I_2(\kappa / 2)}{I_1(\kappa / 2)} \simeq \frac{\kappa}{4}, \quad \frac{I_1(\kappa / 2)}{I_0(\kappa / 2)} \simeq \frac{\kappa}{2},\qquad \text{as }\kappa\to  0,
\end{equation}
and
\begin{equation}\label{eq:quotients2}
\frac{I_2(\kappa / 2)}{I_1(\kappa / 2)} \simeq 1-\frac 3 \kappa, \quad \frac{I_1(\kappa / 2)}{I_0(\kappa / 2)} \simeq 1-\frac 1 \kappa,\qquad\text{as }\kappa\to  +\infty.
\end{equation}
With this result, we compute
\begin{equation}
\begin{aligned}\label{eq:FGhqual1}
&F_1(\kappa) 
\simeq
\displaystyle \frac{2}{\kappa},\
&&
F_2(\kappa)  \simeq 
0, & \qquad \text{as }\kappa\to  0,
\\
&F_1(\kappa) 
\simeq
\displaystyle 1,\
&&
F_2(\kappa)  \simeq 
1, & \text{as }\kappa\to  +\infty.
\end{aligned}
\end{equation}
From \eqref{eq:FGhqual1}, we obtain \eqref{fgh2}. Thus, for large values of the \color{black} order parameter $\kappa$, the differential equation that governs the mean orientation $\mu(t)$ is the same as \eqref{eq:deterministic}, namely, the equation that  governs the orientation of a single cell in the deterministic case. On the other hand, when the \color{black} order parameter \color{black} \color{black}$\kappa$ is very small, that is, the distribution is very broad, $\mu(t)$ approaches $\pi/2$ at a rate that tends to infinity as $\kappa$ tends to zero. In both regimes, as well as the intermediate ones $\mu(t)$, we have
\begin{equation}
\lim_{t\to\infty}\mu(t)=\frac \pi 2,
\end{equation}
with exponential rate, provided that $\mu(0)\neq 0$ and $\mu(0)\neq \pi$.

\subsection*{The evolution of $\kappa(t)$.}\label{subs:kappa}
While the dynamics of the mean orientation $\mu(t)$ is not too dissimilar, at least qualitatively, from that of the angle $\theta(t)$ in the deterministic case, the description of the dynamics of $\kappa(t)$ needs a more elaborate study, as can been be guessed by inspection of the trajectories in Fig.~\ref{fig:phaseportrait}.

\begin{figure}[H]
\centering
\centering
\includegraphics[width=0.6\linewidth]{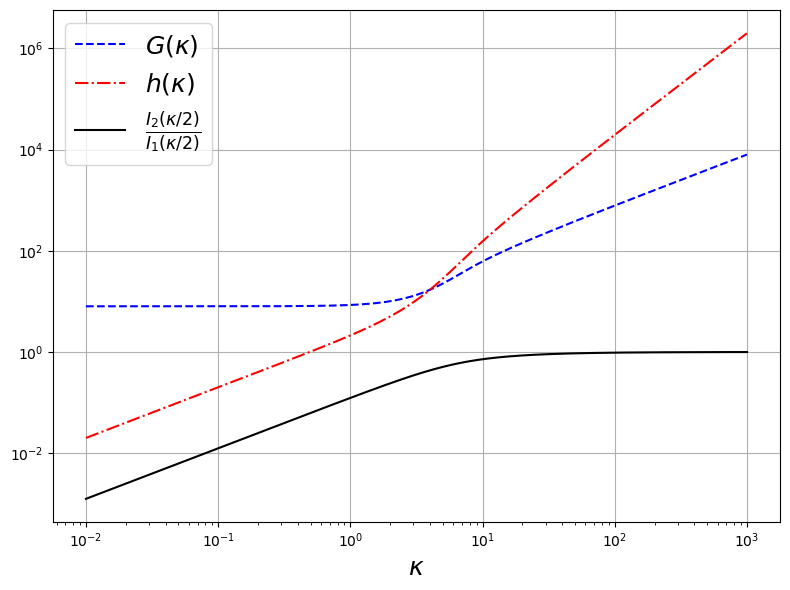}
\caption{(a) Plots of $G(\kappa)$, $h(\kappa)$, and the ratio $\frac{I_2(\kappa/2)}{I_1(\kappa/2)}$.}
\label{fig:plotsGh}
\end{figure}

To better understand the qualitative behaviour of $\kappa(t)$, we rewrite the differential equation that governs the evolution of $\kappa(t)$ as  \begin{equation}\label{ODE-tris}
\dot\kappa=- G(\kappa)\left(\frac{I_2\left(\frac{\kappa}{2}\right)}{I_1\left(\frac{\kappa}{2}\right)}\left(2(\cos 2 \mu)^2-1\right)+k_2\cos 2 \mu\right)- \rho^2 h(\kappa),
\end{equation}
where 
a superior dot denotes the derivative with respect to the 
dimensionless time
\begin{equation}
\overline t=\frac{\eta}{K\varepsilon^2}t,
\end{equation}
and 
\begin{equation}
\rho=\frac\tau \varepsilon.	
\end{equation}\color{black}
The right-hand side of the differential equation \eqref{ODE-tris} is the sum drift term and diffusion term. Their relative importance can be assessed by inspection of the graphs of $G(\kappa)$, $h(\kappa)$ and $\frac{I_2(\kappa/2)}{I_1(\kappa/2)}$ in Fig.~\ref{fig:plotsGh}.
Using the asymptotic properties of the quotient $\frac{I_2(\kappa/2)}{I_1(\kappa/2)}$, which have already been recorded in \eqref{eq:quotients1} and \eqref{eq:quotients2}, we obtain the estimates
\begin{equation}
\begin{aligned}\label{eq:FGhqual}
&G(\kappa) 
\simeq
8,&& h(\kappa) 
\simeq
4\kappa, & \quad \text{as }\kappa &\to 0,
\\
&G(\kappa) 
\simeq
32\kappa,&& h(\kappa) 
\simeq
8\kappa^2, & \quad \text{as }\kappa &\to +\infty.
\end{aligned}
\end{equation}
By \eqref{eq:FGhqual} and from \eqref{eq:quotients1} it follows that
\begin{equation}\label{eq:asymptg} 
\begin{aligned}
& \dot \kappa \simeq -8  k_2 \color{black}\cos 2 \mu,
&& \quad \text{as } \kappa \to 0,\\
& {\dot \kappa} \simeq -16\rho \kappa^2,
&& \quad\text{as } \kappa \to +\infty.
\end{aligned}
\end{equation}  
Thus, for $\kappa$ small, drift is dominant. By contrast, the diffusion term is dominant for large vales of $\kappa$. It should be noted, however, that the effects of drift and diffusion are not always in competition. In fact, while the sign of $h(\kappa)$ is always positive, so that the effect of diffusion results always in a reduction of the order parameter $\kappa$, the drift term is the product of $G(\kappa)$, a positive function, with a polynomial of $\cos 2\mu$, whose sign may be positive or negative. This is evident by looking at the first of \eqref{eq:asymptg}, which shows that  drift promotes --- at least for $\kappa$ small --- the increase of the order parameter $\kappa$ when $\mu$ falls in the ``spinodal region'' $(\frac \pi 2,\frac 3 2 \pi)$. This is also confirmed 
by examining the behaviour of $\dot \kappa$ on the set $\dot\mu=0$, that is, on the vertical lines $\mu=0$, $\mu=\pi/2$, and $\mu=\pi$. Indeed, we have
\begin{equation}\label{eq:signs}
\begin{aligned}
&\dot{\kappa}=G(\kappa)\left(k-\frac{I_2\left(\frac{\kappa}{2}\right)}{I_1\left(\frac{\kappa}{2}\right)}\right)-\rho^2 h(\kappa),&&\qquad \text{for }\mu=\pi/2,\\
&\dot{\kappa}=-G(\kappa)\left(k+\frac{I_2\left(\frac{\kappa}{2}\right)}{I_1\left(\frac{\kappa}{2}\right)}\right)-\rho^2 h(\kappa),&&\qquad \text{for }\mu=0,\pi.
\end{aligned}
\end{equation}
Clearly, since $k\ge 1$, and since $0<\frac{I_2\left(\frac{\kappa}{2}\right)}{I_1\left(\frac{\kappa}{2}\right)}<1$ (recall Fig.~\ref{fig:plotsGh}) the drift term in the first of \eqref{eq:signs} is always positive, while it is always negative in the second of \eqref{eq:signs}. Hence, drift competes with diffusion when $\mu=\frac \pi 2$ (where the energy $\mathscr U(\mu)$ has its minumum), and cooperates with diffusion when $\mu=0$ or $\mu=\pi$ (where $\mathscr U(\mu)$ has its maximum). This observation is in accordance with what emerges from the phase-plane trajectories in Fig.~\ref{fig:phaseportrait}(a).

We note from the first of \eqref{eq:signs} that there exists a unique point $\kappa_*$ on the vertical line $\mu=\pi/2$ such that $\dot\kappa=0$. This point solves the equation (\emph{cf.} \eqref{eq:kappastar2})
\begin{equation}\label{eq:kappastar}
k b(\kappa_*)-\rho^2c(\kappa_*)-1=0,
\end{equation}
where 
\begin{equation}
 b(\kappa)=\frac{I_1\left(\frac{\kappa}{2}\right)}{I_2\left(\frac{\kappa}{2}\right)},\qquad c(\kappa)=\frac \kappa 2 \frac{I_1\left(\frac{\kappa}{2}\right)}{I_2\left(\frac{\kappa}{2}\right)}.
\end{equation}
This leads to \eqref{eq:kappastar2}.
\begin{figure}[H]
\centering
\begin{subfigure}[t]{0.48\linewidth}
\centering
\includegraphics[width=\linewidth]{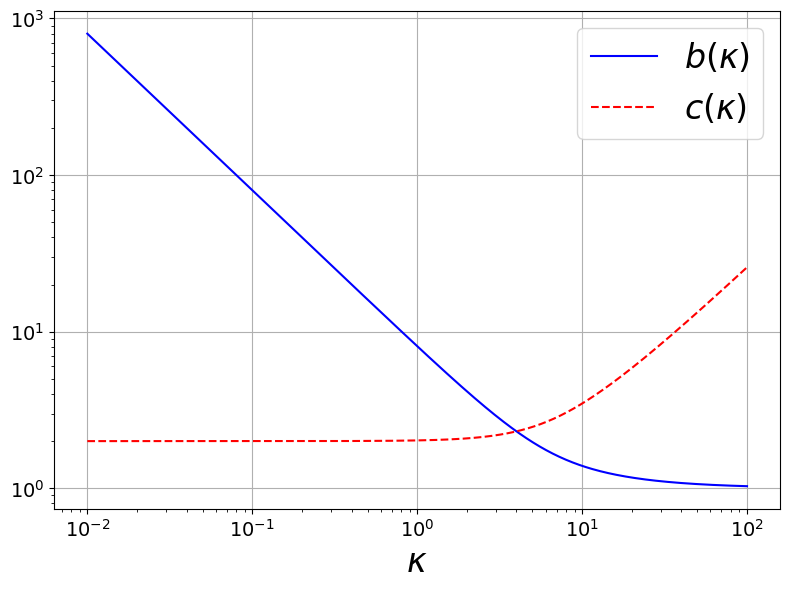}
\caption{Log-log plots of $b(\kappa)$ and $c(\kappa)$.}
\label{fig:coeff-bc}
\end{subfigure}
\hfill
\begin{subfigure}[t]{0.48\linewidth}
\centering
\includegraphics[width=\linewidth]{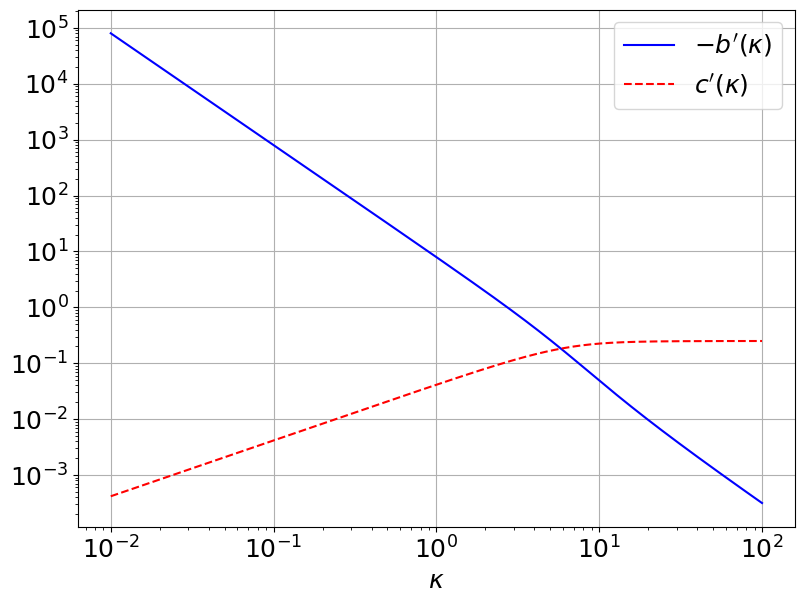}
\caption{Log-log plots of $-b'(\kappa)$ and $c'(\kappa)$.}
\label{fig:coeff-derivatives}
\end{subfigure}
\caption{Properties of $b(\kappa)$ and $c(\kappa)$.}
\label{fig:coefficients}
\end{figure}

The functions $b(\kappa)$ and $c(\kappa)$, whose graphs are given in Fig.~\ref{fig:coefficients}, have the following properties. They are both positive and convex, furthermore $b(\kappa)$ is decreasing and $c(\kappa)$ is increasing. Thus, the left-hand side of \eqref{eq:kappastar} is a decreasing with $\kappa$. Moreover, it is positive for $\kappa$ small and negative for $\kappa$ large. Hence, there exists a unique $\kappa_*=\kappa_*(\rho,\kappa)$ such that $\dot\kappa=0$ on the vertical line $\mu=\pi/2$. This is the unique fixed point of the dynamical system \eqref{ODE-bis}. 
Further, we compute
\begin{equation}\label{eq:monot2bis}
\frac{\partial\kappa_*}{\partial k}=\frac{b(\kappa_*)}{\rho^2 c'(\kappa_*)-kb'(\kappa_*)}>0,\qquad \frac{\partial\kappa_*}{\partial 
\rho^2}=-\frac{c(\kappa_*)}{\rho^2 c'(\kappa_*)-k b'(\kappa_*)}<0.
\end{equation}
In particular, $\kappa_*$ is a decreasing function of $\rho$. This, is consistent with the observation that increasing the strain amplitude increases the order parameter, and hence decreases the spreading of the probability distribution. Furthermore, we notice that 
\begin{equation}
b(\kappa)\simeq \frac{4}{\kappa},\qquad c(\kappa)\simeq 2,\qquad \text{as }\kappa\to 0,
\end{equation}
and that, moreover, 
\begin{equation}\label{eq:limit3}
b(\kappa)\simeq 1,\qquad c(\kappa)\simeq \frac \kappa 4 ,
\qquad 	\text{as }\kappa\to\infty.
\end{equation}
where oblique arrows denote monotone convergence.
This observation allows us to obtain the asymptotic estimates \eqref{eq:kappastar3}.

\subsection*{The set $\dot\kappa=0$.}
We next illustrate additional properties of the set ${\bf\dot\kappa=0}$. Fig.~\ref{fig:phaseportrait}(b) suggests that the set $\dot\kappa>0$ is a bounded region on the phase plane. To substantiate this hypothesis we now characterise the set of pairs $(\mu,\kappa)$ such that $\dot\kappa=0$. 

\paragraph{The polynomial $\bf P_\kappa({\bf \it x})$.}
By \eqref{eq:asymptg} we already know that the set $\dot\kappa=0$ includes two mirror-symmetric points with coordinates $(\mu,\kappa)=(\frac{\pi}{4},0)$ and $(\mu,\kappa) = (\frac{3}{4}\pi,0)$ on the horizontal line $\kappa = 0$, and that, moreover, $\dot\mu>0$ in the segment $\mu\in (\frac\pi 2,\frac 34\pi)$ \color{black}along \color{black}that line. This finding is consistent with Fig.~\ref{fig:phaseportrait}(b). 

To study the set $\dot\kappa=0$ for $\kappa>0$ we shall analyze the sign of $\dot\kappa$ along horizontal slices of the phase plane. Our motivation for proceeding in this manner is that, as can be seen from \eqref{ODE-tris}, $\dot\kappa$ is a polynomial of degree two with respect to $\cos 2\mu$ when $\kappa>0$. In particular, we can rewrite \eqref{ODE-tris} as
\begin{equation}
\dot\kappa = - G(\kappa) \frac{I_2\left(\frac{\kappa}{2}\right)}{I_1\left(\frac{\kappa}{2}\right)} P_\kappa(\cos2\mu),
\end{equation}
where $P_\kappa(x)$ is the second-order polynomial defined by
\begin{equation}
P_\kappa(x)=2x^2+k_2 b(\kappa)x+\rho^2c(\kappa)-1,
\end{equation}

Hence, to determine the rest of the set $\dot\kappa=0$ it suffices for us to look for the pairs $(\mu,\kappa)$ with $0<\mu<\pi$ and $\kappa>0$ such that
\begin{equation}\label{eq:P}
P_\kappa(\cos 2\mu)=0,
\end{equation}
which is equivalent to finding the roots of $P_\kappa(x)$ in the interval $[-1,+1]$.

\begin{figure}[H]
\centering
\includegraphics[width=0.5\linewidth]{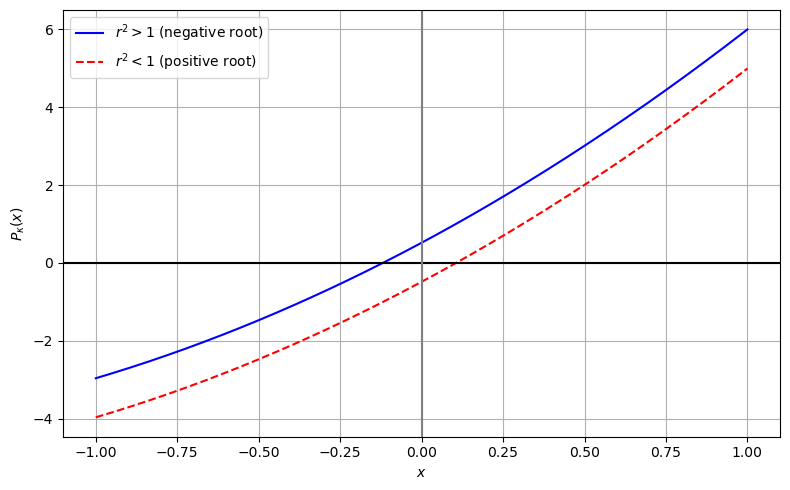}
\caption{Illustrative plot of $P_\kappa(x)$ for $\kappa\ll 1$.}\label{fig:parabolas}
\label{fig:parabola}
\end{figure}

\paragraph{The roots $x_1(\kappa)$ and $x_2(\kappa)$.} 
The polynomial $P_\kappa(x)$ is convex and its derivative 
\begin{equation}\label{eq:deriv2}
P'_{\kappa}(x)=4x+k_2 b(\kappa),
\end{equation}
vanishes at the point 
\begin{equation}\label{eq:minP}
x_{\rm min}(\kappa)=-\frac {k_2}4 b(\kappa)<-\frac {k_2}4.
\end{equation}
In this point, $P_\kappa(x)$ has its minimum, given by
\begin{equation}
P_{\min}(\kappa)=\min_x P_\kappa(x)=\rho^2c(\kappa)-\frac {k_2^2}8 b^2(\kappa)-1.
\end{equation}
Since $b(\kappa)$ is decreasing and $c(\kappa)$ is increasing,  $P_{\rm min}(\kappa)$ is increasing with $\kappa$. Moreover, since $b(\kappa)\to +\infty$ as $\kappa\to 0$, and $c(\kappa)\to +\infty$ as $\kappa\to+\infty$, we have that $P_{\min}(\kappa)\to-\infty$ as $\kappa\to 0$, and that $P_{\rm min}(\kappa)\to +\infty$ as $\kappa\to+\infty$. This implies that there exists $\bar\kappa$ such that
\begin{equation}
P_{\rm min}(\bar\kappa)=0.
\end{equation}
For $\kappa>\bar\kappa$, the polynomial $P_\kappa(x)$ has no real roots, hence the equation \eqref{eq:P} has no solution. This implies that the set $\dot\kappa=0$ is bounded from above, with $\bar\kappa$ being an upper bound for the vertical coordinate $\kappa_{\rm max}$ of its uppermost points. For $\kappa<\bar\kappa$ the polynomial $P_\kappa(x)$ has two distinct real roots $x_\alpha(\kappa)$, $\alpha=1,2$ satisfying
\begin{equation}
P_\kappa(x_\alpha(\kappa)) = 0.
\end{equation}
We label these roots so that
\begin{equation}
x_1(\kappa)< x_2(\kappa),
\end{equation}
and we observe for later that, since $P_\kappa(x)$ is convex,
\begin{equation}
P'_\kappa(x_1(\kappa))<0,\qquad P'_\kappa(x_2(\kappa))>0.
\end{equation}
If at least one of the roots $x_\alpha(x)$ is in the interval $[-1,+1]$ then the equation \eqref{eq:P} has two or more solutions. We define
\begin{equation}\label{eq:defmualpha}
\mu_\alpha(\kappa)=\frac 1 2\arccos(x_\alpha(\kappa)),\qquad \text{if } x_\alpha(\kappa)\in[-1,+1], 
\end{equation}
so that $\mu_\alpha(\kappa)$ and $\pi-\mu_\alpha(\kappa)$ are solutions of \eqref{eq:P}.
\paragraph{Behaviour of the roots $x_1(\kappa)$ and $x_2(\kappa)$ for small $\kappa$.}
It has been observed that the minimum of the polynomial $P_\kappa(x)$ is located always at $x$ negative (\emph{cf.} \eqref{eq:minP}). This implies that the smallest root of $P_\kappa(x)$ is negative as well:
\begin{equation}\label{eq:x1lessthan0}
x_1(\kappa)<0\quad\text{for all }\kappa>0,
\end{equation}
and that
\begin{equation}
x_1(\kappa)\to-\infty,\qquad \text{as }\kappa\to 0.
\end{equation}
Furthermore, for $\kappa$ small, the graph of  $P_\kappa(x)$ becomes a steep line having positive slope passing though the point of coordinates $x=0$ and $y=P_\kappa(0)=\rho^2 c(0)-1= \rho^2-1$. This implies that
\begin{itemize}
\item as $\kappa\to 0$, the largest root $x_2(\kappa)$ of $P_\kappa(x)$ converges to 0 from below if $\rho^2>1$ and from above if $\rho^2<1$,
\end{itemize}
as illustrated Fig.~\ref{fig:parabolas}.

\paragraph{Behaviour of the roots $x_\alpha(\kappa)$ as $\kappa$ increases.} 
We now characterize more in detail how the locations of the roots $x_\alpha(\kappa)$ change as $\kappa$ increases from $0$ to $\infty$. To begin with, we differentiate the equation $P_\kappa(x_\alpha(\kappa)) = 0$ with respect to $\kappa$ to obtain
\begin{equation}\label{eq:derivative}
\frac{d x_\alpha(\kappa)}{d \kappa}=-\frac{\rho^2c'(\kappa)+k_2b'(\kappa)x_\alpha(\kappa)}{P'_\kappa(x_\alpha(\kappa))},\qquad \alpha=1,2,
\end{equation}
Since $b'(\kappa)<0$, $c'(\kappa)>0$, and $x_1(\kappa)<0$ (see \eqref{eq:x1lessthan0}), and since $P'_{\kappa}(x_1(\kappa))<0$, we conclude that the root $x_1(\kappa)$ is increasing:
\begin{equation}
\frac{d x_1(\kappa)}{d \kappa}>0.
\end{equation}
By the same argument we see that
\begin{equation}
x_2(\kappa)<0\quad\Rightarrow \frac{d x_2(\kappa)}{d \kappa}<0.
\end{equation}
From what has been observed on the behaviour of the root $x_2(\kappa)$ for $\kappa\to 0$, it follows that the root $x_2(\kappa)$ is decreasing if $\rho^2>1$; and non-monotone  if $\rho^2<1$. 

\begin{figure}[H]
\centering
\includegraphics[width=0.6\linewidth]{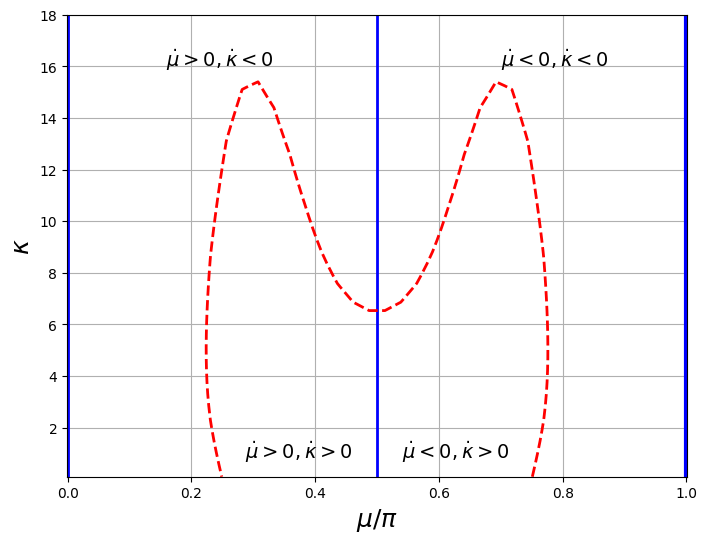}
\caption{Zero-level sets of $\dot\mu$ and $\dot\kappa$ in the case $\rho^2<1$. The root $x_2(\kappa)$ is nonmonotone. Accordingly, the corresponding value $\mu_2(\kappa)$ from \eqref{eq:defmualpha} is nonmonotone as well.}
\label{fig:phaseportrait3}
\end{figure}

In particular, for $\rho^2<1$ there exists a critical value $\hat\kappa$ such that $x_2(\kappa)$ is increasing for $0<\kappa<\hat\kappa$, and decreasing for $\kappa>\hat\kappa$. To verify this assertion, we first differentiate \eqref{eq:derivative} with respect to \(\kappa\) to obtain:
\begin{equation}
\frac{d^2 x_\alpha}{d\kappa^2} = 
- \frac{
\left( \rho^2 c''(\kappa) + k b''(\kappa) x_2(\kappa) + k b'(\kappa) \frac{d x_\alpha}{d\kappa} \right)
}{
P'_\kappa(x_2) 
}
- \frac{d x_\alpha}{d\kappa}\frac{
P''_\kappa(x_2)
}{
P'_\kappa(x_2) 
}.
\end{equation}
Moreover, whenever \( x_2(\kappa) > 0 \) and \( \frac{d x_2}{d\kappa} = 0 \),
\begin{equation}
\left. \frac{d^2 x_2}{d\kappa^2} \right|_{\frac{dx_2}{d\kappa}=0}
= - \frac{ \rho^2 c''(\kappa) + k b''(\kappa) x_2(\kappa) }{ P'_\kappa(x_2) }<0,
\end{equation}
since $b(\kappa)$ and $c(\kappa)$ are convex, and \( P'_\kappa(x_2) > 0 \). Thus, even if $x_2(\kappa)$ is not monotone, once it starts decreasing, it continues to decrease. Eventually, since $x_1(\bar\kappa)=x_2(\bar\kappa)$, and since $x_1(\kappa)$ can never be positive, the root $x_2(\kappa)$ is negative for $\kappa$ sufficiently large. An illustrative plot of the sets $\dot\mu=0$ and $\dot\kappa=0$ is shown in Fig.~\ref{fig:phaseportrait3}
\begin{remark}[Non-monotone behaviour of $\kappa(t)$]\label{rem:nonmonotone}
	We observe that in Fig.~\ref{fig:phaseportrait3} the set $\dot\kappa>0$ has two lobes that extend well above the vertical line that crosses the stationary point $\kappa_*$. If the initial state of the dynamical system is in one of the two lobes, then the concentration parameter $\kappa(t)$ first increases, and then decreases. This interesting behavior is observed in Fig.~\ref{fig:kappa4}.\color{black}
	\end{remark}

\paragraph{The intersection of $\dot\kappa=0$ with the central vertical axis $\mu=\pi/2$.}
From the above analysis, we see that two scenarios are possible: either the smallest root $x_1(\kappa)$ of the polynomial $P_\kappa(x)$ remains smaller than $-1$ for all $\kappa$, in which case the equation $P_\kappa(\cos 2\mu) = 0$ admits at most two mirror-symmetric solutions for each $\kappa$; or there exists a value $\kappa^*$ such that $x_1(\kappa^*) = -1$, leading to the appearance of four roots solutions for some values of $\kappa$.

To discriminate between these cases, we seek values of $\kappa$ such that one of the roots, either $x_1(\kappa)$ or $x_2(\kappa)$ hits $x=-1$. To this aim, we solve the equation 
\begin{equation}\label{eq:critical}
0=P_{\kappa}(-1)=1-k_2b(\kappa)+\rho^2 c(\kappa)
\end{equation}
for the unknown $\kappa$. The existence of a unique solution $\kappa_* = \kappa_*(k_2, \rho^2)$ has already been discussed, along with the observation that such solution determines the unique fixed point of the dynamical system \eqref{ODE-bis}.

\begin{figure}[H]
\centering
\includegraphics[width=0.5\linewidth]{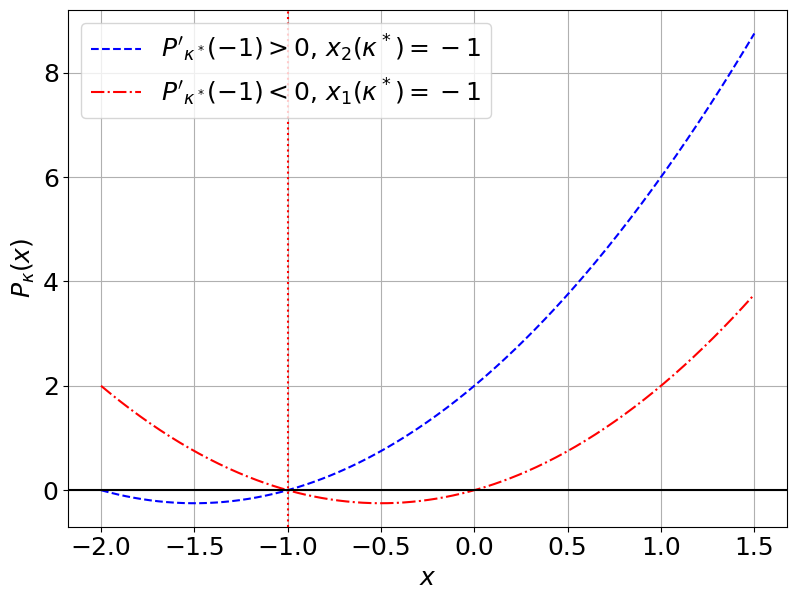}
\caption{
Example of two possible configurations when the polynomial $P_{\kappa}(x)$ has a root at $x = -1$. In the first case (blue dashed curve), the root corresponds to the smallest solution $x_1(\kappa)$ and the derivative $P'_{\kappa}(-1)$ is negative. In the second case (red dashed-dotted curve), the root corresponds to the largest solution $x_2(\kappa)$ and $P'_{\kappa}(-1)$ is positive. The dashed red line highlights the location $x = -1$.
}\label{fig:derivativePkappa}
\end{figure}

\begin{figure}[H]
\centering
\includegraphics[width=0.6\linewidth]{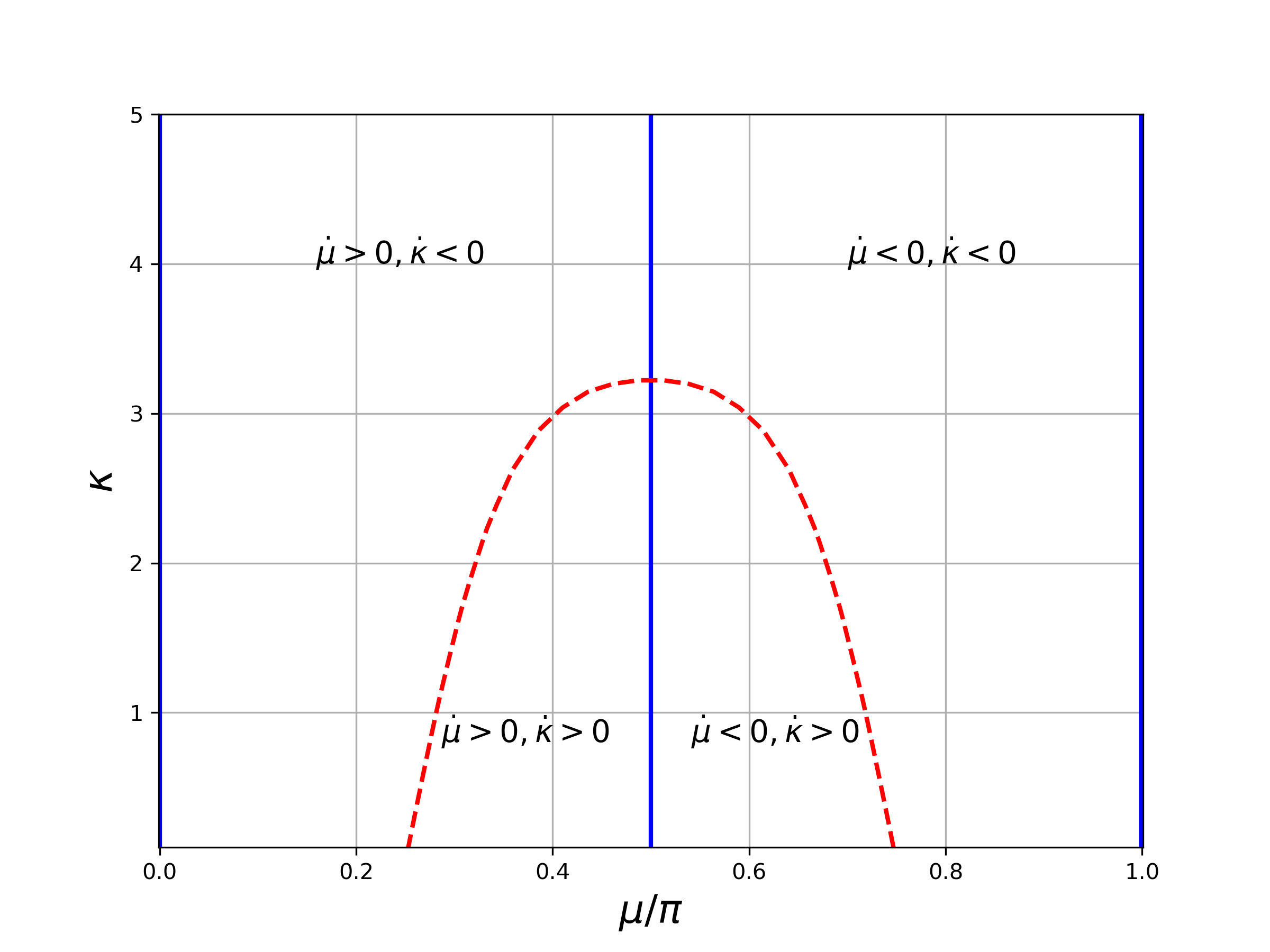}
\caption{Zero-level sets of $\dot\mu$ and $\dot\kappa$ in the case $P'_{\kappa_*}(-1)>0$ (dashed blue line in Fig.~\ref{fig:derivativePkappa}). In this case the smallest root never hits the interval $[-1,+1]$, thus for fixed $\kappa$ there are only two values of $\mu$ such that $\dot\kappa=0$. As a result, the set $\dot\kappa>0$, which lies below the dashed curve, is convex. If $P'_{\kappa_*}(-1)<0$, the set is non-convex (see Fig.~\ref{fig:phaseportrait}(b).) This curve has been computed taking $k=2$ and $\rho^2=2$.}
\label{fig:phaseportrait2}
\end{figure}

The root at $x = -1$ may correspond either to $x_1(\kappa^*)$ or to $x_2(\kappa^*)$. To distinguish between these two cases, we examine the sign of the derivative
\begin{equation}\label{eq:deriv}
P'_{\kappa^*}(-1)=k_2 b(\kappa^*)-4,
\end{equation}
since $P'_{\kappa^*}(-1)<0$ if $x_1(\kappa^*)=-1$, and  $x_2(\kappa^*)=-1$ otherwise, as shown in the illustrative plot in Fig.~\ref{fig:derivativePkappa}.
If $P_{\kappa_*}(-1)<0$ (so that $x_1(\kappa_*)=-1$), then the polynomial $P_{\kappa_*}(x)$ has both roots $x_1(\kappa^*)$ and $x_2(\kappa^*)$ in the interval $[-1,+1]$ and as $\kappa$ increases further, $x_1(\kappa^*)$ increases and $x_2(\kappa^*)$ decreases, until they are the same. In this case, the region $\dot\kappa>0$, which is below the set $\dot\kappa=0$, is nonconvex, as in Fig.~\ref{fig:phaseportrait}(b). If, instead, $P_{\kappa_*}(-1)>0$ (so that $x_2(\kappa_*)=-1$), then the polynomial $P_{\kappa_*}(x)$ has only the root $x_2(\kappa_*)$  in the interval $[-1,+1]$. As $\kappa$ increases further, the root $x_2(\kappa_*)$ leaves the interval $[-1,+1]$, and therefore equation $P_{\kappa}(\cos 2\mu)$ does not have a solution $\mu\in[0,\pi]$ for $\kappa>\kappa_*$. In the latter case, the region $\dot\kappa>0$, which is below the set $\dot\kappa=0$, is convex, as shown in the illustrative example of Fig.~\ref{fig:phaseportrait2}
\paragraph{Convexity of the set $\bf \dot\kappa>0$.} Whether the set $\dot\kappa>0$ is convex or not, this depends on the sign of $P'_{\kappa_*}(-1)$ in \eqref{eq:deriv}. Since $\kappa_*$ is a function of $k_2$ and $\rho^2$, the sign of $P'_{\kappa_*}(-1)$ ultimately depends on $k_2$ and $\rho^2$. To better understand such a dependence, we study the partial derivatives of $P'_{\kappa_*}(-1)$ with respect to $k_2$ and $\rho^2$. With this goal in mind, we first compute
\begin{equation}\label{eq:monot2}
\frac{\partial\kappa_*}{\partial k_2}=\frac{b(\kappa_*)}{\rho^2c'(\kappa_*)-k_2b'(\kappa_*)}>0,\qquad \frac{\partial\kappa_*}{\partial \rho^2}=-\frac{c(\kappa_*)}{\rho^2c'(\kappa_*)-k_2b'(\kappa_*)}<0.
\end{equation}
Therefore,
\begin{equation}
	\frac{\partial P'_{\kappa_*}(-1)}{\partial \rho^2}	=k_2 b'(\kappa_*)\frac{\partial\kappa_*}{\partial \rho^2}>0.
\end{equation}
The partial derivative with respect to $k_2$ can be computed by noting that, by \eqref{eq:critical} and  \eqref{eq:deriv},
\begin{equation}
P_{\kappa^*}^{\prime}(-1)=\rho^2 c\left(\kappa^*\right)-4.	
\end{equation}
Therefore,
\begin{equation}
\frac{\partial P'_{\kappa_*}(-1)}{\partial k_2} =\rho^2 c'(\kappa_*)\frac{\partial\kappa_*}{\partial k_2}>0.
\end{equation}
Since $b(\kappa)$ is monotone, it follows \eqref{eq:deriv} and from \eqref{eq:monot2} that  $P_{\kappa_*}'(-1)$ depends monotonically both on $k_2$ and $\rho^2$. This means that the plane $k$--$\rho^2$ is divided in two regions where $P_{\kappa_*}'(-1)>0$ and $\dot\kappa>0$ is convex, and $P_{\kappa_*}'(-1)<0$ where $\dot\kappa>0$ is nonconvex. These regions are separated by the curve defined by the equation $P_{\kappa_*}'(-1)=0$. This curve is described by the parametric equations
\begin{equation}
\begin{aligned}
&k_2(\kappa^*)=\frac 4 {b(\kappa^*)},\\
&\rho^2(\kappa^*)=\frac 4 {c(\kappa^*)}.\\
\end{aligned}
\end{equation}

\begin{figure}[H]
\centering
\includegraphics[width=0.6\linewidth]{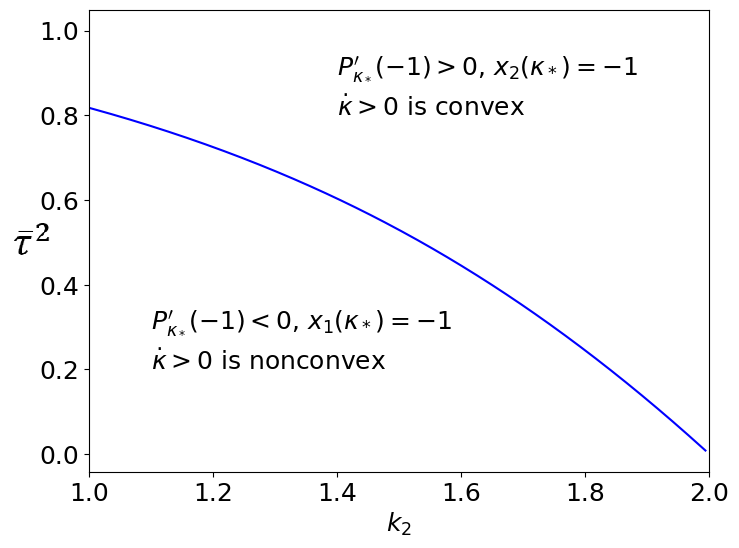}
\caption{Curve in the $(k_2,\rho^2)$-plane corresponding to the condition $P_{\kappa^*}(-1) = 0$.}
\label{fig:k2-vs-r2k1}
\end{figure}
\end{document}